# Rotational ultrasound and photoacoustic tomography of the human body

Yang Zhang[1,2†], Shuai Na[1,7†], Jonathan J. Russin[3,4,5†], Karteekeya Sastry[1,6], Li Lin[1,8], Junfu Zheng[1], Yilin Luo[1], Xin Tong[1], Yujin An[1], Peng Hu[1], Konstantin Maslov[1], Tze-Woei Tan[9], Charles Y. Liu[3,4,5, *], Lihong V. Wang[1,6,*]

[1] Caltech Optical Imaging Laboratory, Andrew and Peggy Cherng Department of Medical Engineering, California Institute of Technology, 1200 East California Boulevard, Pasadena, CA 91125, USA.

[2] School of Biomedical Engineering, Tsinghua University, Beijing, 100084, China.

[3] Department of Neurological Surgery, Keck School of Medicine, University of Southern California, Los Angeles, CA 90033, USA.

[4] Neurorestoration Center, Keck School of Medicine, University of Southern California, Los Angeles, CA 90033, USA.

[5] Rancho Los Amigos National Rehabilitation Center, Downey, CA 90242, USA.

[6] Caltech Optical Imaging Laboratory, Department of Electrical Engineering, California Institute of Technology, 1200 East California Boulevard, Pasadena, CA 91125, USA.

[7] Present address: China National Biomedical Imaging Center, College of Future Technology, Peking University, Beijing, 100871, China.

[8] Present address: College of Biomedical Engineering and Instrument Science, Zhejiang University, Hangzhou, China.

[9] Keck School of Medicine, University of Southern California, Los Angeles, CA 90033, USA.

[†]These authors contributed equally to this work: Yang Zhang, Shuai Na, Jonathan J. Russin.

*Corresponding author. C. Y. Liu (cliu@usc.edu), L. V. Wang (LVW@caltech.edu).

## Abstract

Imaging the human body's morphological and angiographic information is essential for diagnosing, monitoring, and treating medical conditions. Ultrasonography performs the morphological assessment of the soft tissue based on acoustic impedance variations, whereas photoacoustic tomography (PAT) can visualize blood vessels based on intrinsic hemoglobin absorption. Three-dimensional (3D) panoramic imaging of the vasculature is generally not practical in conventional ultrasonography with limited field-of-view (FOV) probes, and PAT does not provide sufficient scattering-based soft tissue morphological contrast. Complementing each other, fast panoramic rotational ultrasound tomography (RUST) and PAT are integrated for hybrid rotational ultrasound and photoacoustic tomography (RUS-PAT), which obtains 3D ultrasound structural and PAT angiographic images of the human body quasi-simultaneously. The RUST functionality is achieved in a cost-effective manner using a single-element ultrasonic transducer for ultrasound transmission and rotating arc-shaped arrays for 3D panoramic detection. RUST is superior to conventional ultrasonography, which either has a limited FOV with a linear array or is high-cost with a hemispherical array that requires both transmission and receiving. By switching the acoustic source to a light source, the system is conveniently converted to PAT mode to acquire angiographic images in the same region. Using RUS-PAT, we have successfully imaged the human head, breast, hand, and foot with a 10 cm diameter FOV, submillimeter isotropic resolution, and 10 s imaging time for each modality. The 3D RUS-PAT is a powerful tool for high-speed, 3D, dual-contrast imaging of the human body with potential for rapid clinical translation.

## Introduction

Imaging techniques for visualizing the human body's morphological and angiographic information, such as magnetic resonance imaging (MRI)[1] and *X*-ray computed tomography (CT)[2], have paved the way for medical diagnoses and physiological studies. MRI detects signals mainly from protons found in water that makes up living tissues[1], and it is routinely used in clinical applications for structural imaging throughout the body, and functional imaging of the brain. CT measures the *X*-ray attenuation of different tissues and bones inside the body[2], and it is used to obtain anatomical information about the human body for medical diagnoses with clear advantages to MRI in diagnosing bone disorders and acute hemorrhage. Magnetic resonance angiography (MRA) can image the body's blood vessels[3]. With the use of contrast agents, CT can image the angiographic information of the human body[4]. However, MRI is generally high-cost, CT involves exposure to ionizing radiation, and the contrast agents have side effects.

In addition to methods based on *X*-ray and magnetic fields, acoustic and optical techniques have been developed for imaging the human body in a safe and relatively cost-effective manner. Due to its low cost and portability, ultrasonography has been widely used in clinical applications to image the heart, breast, and viscera based on acoustic impedance variations[5]. However, current clinical ultrasonography is generally based on handheld phased/linear arrays[6] which have a limited field-of-view (FOV)[7]. For instance, these arrays have been successfully employed to scan and generate full-volume breast images[8,9]; however, the scanning of these arrays faces the challenges of non-isotropic spatial resolutions. Thus, achieving a three-dimensional (3D) panoramic FOV often requires a hemispherical array configuration and many pulsers for transmission, which is not cost-effective[10,11]. In addition to scanning a linear array and using a dense hemispherical array, several other techniques exist for 3D ultrasound imaging, including the utilization of single-element transducers[11], row-column arrays[12,13], and matrix array probes[14]. Although the use of a single ultrasound sensor with a plastic aperture mask and compressed sensing has demonstrated potential in obtaining 3D images of simple structures like letters[11], its practical application for in vivo cases remains unexplored. Row-column arrays have been investigated for 3D anatomical and functional imaging, but their limited array elements result in increased image artifacts and compromised spatial resolutions due to line-based focusing[12,13]. Matrix array probes offer full control in transmission and reception mode, enabling fully focused ultrasound images for applications such as brain imaging within a 1 cm diameter FOV[14]. However, the quadratic increase in the number of array elements/channels with larger apertures poses practical challenges and is not ideal for low-cost applications. Consequently, the quest for achieving high-speed, large field-of-view, and submillimeter isotropic resolution in 3D ultrasound imaging remains an ongoing challenge in the field.

In addition, although ultrafast Doppler ultrasound has been used to image the vasculature in the brain, breast, and abdomen based on the Doppler effect[15,16], it is mainly limited to the two-dimensional (2D) images. Furthermore, the application of ultrafast ultrasound localization microscopy, which involves the use of injected microbubbles, facilitates transcranial imaging of deep vasculature in the adult human brain with microscopic resolution[17]. This technique also enables the imaging of myocardial vasculature in patients through transthoracic approaches[18].

However, it is important to note that the minimally invasive requirement for microbubble injection poses limitations on its wider application. Also, it will require a large number of 2D or 3D array elements and thousands of volumes of data to generate a 3D angiographic image, which is currently not practical for a large FOV (i.e., 10 cm diameter) application. Photoacoustic tomography (PAT) is based on the photoacoustic effect and combines optical absorption contrast with the high spatial resolution of ultrasound[19]. The ability to directly visualize blood vessels and to detect changes in the hemoglobin content enables the acquisition of angiographic images and extraction of functional information[20]. Although PAT has the ability to achieve high-speed large FOV 3D angiographic images[21,22], it generally does not provide sufficient scattering-based morphological contrast. One straightforward approach to combine ultrasound and photoacoustic tomography is by utilizing the same transducer array[23,24]. For instance, incorporating a light source into an existing handheld ultrasound probe enables dual-modality photoacoustic and ultrasound imaging. However, linear array-based photoacoustic imaging often faces limitations related to the limited view angle. Another approach involves the integration of a hemispherical detector array-based photoacoustic tomography system with linear array-based B-mode ultrasound imaging to obtain dual-modality images[25]. However, the ultrasound images generated by this method are impacted by the presence of additional hardware components and non-isotropic spatial resolution. Overall, imaging the human body's morphological and angiographic information in a high-speed, large 3D FOV, cost-effective and safe way remains challenging.

To address this challenge, we present rotational ultrasound tomography (RUST) and its combination with PAT as hybrid rotational ultrasound and photoacoustic tomography (RUS-PAT). RUST uses a single source for ultrasound transmission and a rotating array module for detection to obtain three-dimensional (3D) structures of the human body with a large FOV (~ 10 cm diameter) cost-effectively. In detail, we generate wide-field acoustic waves using a single-element ultrasonic transducer and co-axially rotate the arc-shaped detection ultrasonic transducer array to achieve 3D panoramic hemispherical ultrasonic detection. A 3D volume is obtained by a voxel-based image reconstruction algorithm that considers all the detection positions after the co-axial azimuthal scan of the detection array module. A significant advantage of RUST is that it is fully compatible with PAT. By adding light illumination to the imaging object, we have integrated RUST with PAT as a hybrid RUS-PAT system, in which PAT shares the same detection mechanism with RUST. The presented hybrid tomography offers four advantages. (1) RUST outperforms the current phased/linear array-based ultrasonography by providing 3D panoramic ultrasonic detection. (2) RUST is highly compatible with the PAT system and does not require significant hardware modification. In contrast, adding conventional ultrasound tomography to a PAT system requires adding many pulsers and switches because of PAT's detection-only front-end circuit. (3) RUS-PAT is cost-effective due to the use of a single-element ultrasonic transducer for transmission in RUST and arc-shaped arrays for detection instead of a high-cost dense hemispherical array in either RUST or PAT mode. (4) The hybrid tomography acquires both the ultrasound structural and PAT angiographic volumes of the human body at a high speed, with a large FOV, and in a safe way, which may offer the opportunity for early detection and frequent tracking of diseases including cancer[26,27].

In this study, we have used RUS-PAT to image four disparate anatomical sites in the human body — the head, breast, hand, and foot with a 10-cm-diameter FOV, submillimeter spatial resolution, and 10 s imaging time for each modality. The first known 3D dual-contrast head images of a hemicraniectomy patient, presented here, reveal the boundaries of the scalp and cortical region by RUST, and the vasculature in the head by PAT. These images demonstrate the potential of RUS-PAT for the evaluation of head injuries, the diagnosis of brain disease, and the study of brain function. The dual-modality breast images in healthy subjects demonstrate the capability of RUST to visualize the boundary and the internal structure of the whole breast and PAT to capture the blood vessel distribution inside the breast, which potentially benefits breast cancer diagnosis. The hand images show the boundaries, muscular tissue structure, and blood vessel distribution in the fingers, palm, and back of the hand, and exemplifies the potential to play a key role in the evaluation and treatment of extremity injuries and peripheral vascular disease. The foot images demonstrate the system's ability to detect abnormalities in the foot, particularly in patients with diabetic foot ulcers. By providing detailed visualization of vascular abnormalities and tissue structures, RUS-PAT offers valuable insights into the underlying pathophysiology of diabetic foot ulcers, paving the way for more effective management strategies and personalized care for diabetic patients.

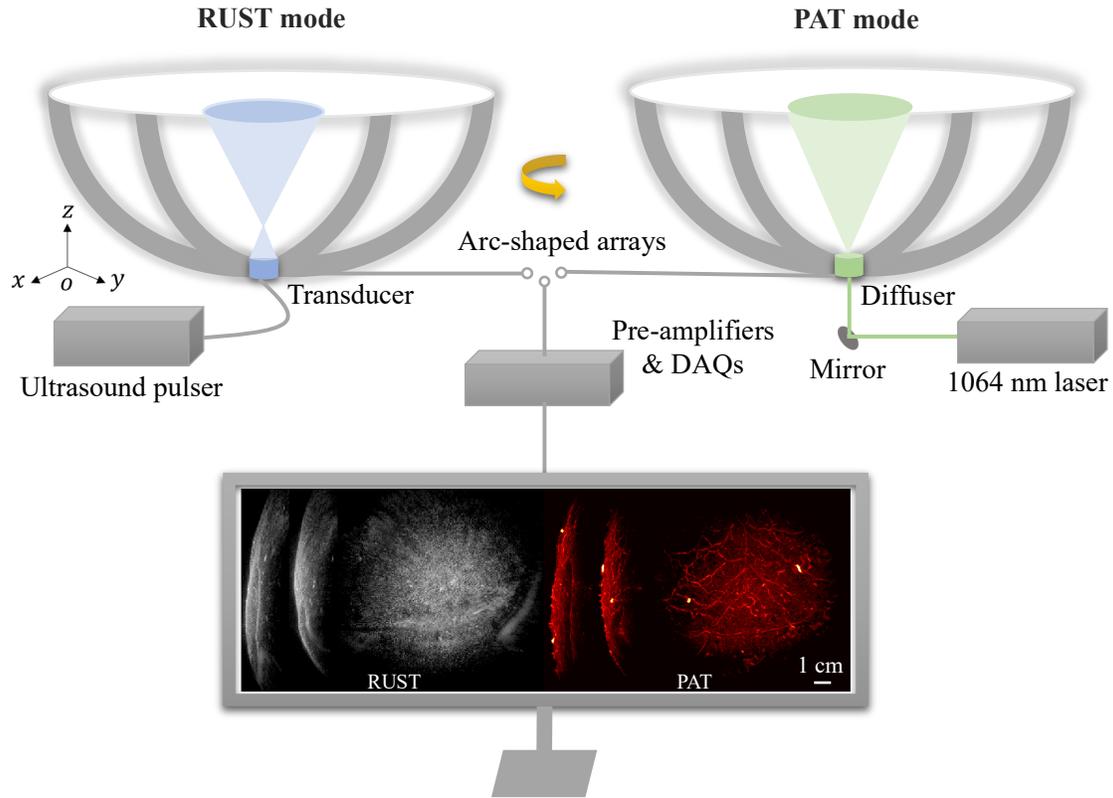

**Fig. 1 | Schematic of RUS-PAT.** RUS-PAT can be operated in the RUST mode and the PAT mode. In the RUST mode, a single-element ultrasonic transducer is mounted co-axially with a rotational stepper motor at the intersection of the arc-shaped detection arrays. In the PAT implementation, an engineered diffuser with a lens tube is installed at the intersection of the arc-shaped detection arrays instead of the transducer in RUST. RUST and PAT share the same detection module, which consists of arc-shaped arrays, pre-amplifiers, DAQs, and a processing system. DAQs: Data acquisition modules; RUST: Rotational ultrasound tomography; PAT: Photoacoustic tomography; RUS-PAT: Rotational ultrasound and photoacoustic tomography.

## Results

**RUS-PAT implementation**. In this article, we present a fast 3D ultrasound tomography method RUST, and its combination with PAT as a hybrid RUS-PAT system to image the human body's morphological and angiographic information in a high-speed, large FOV, cost-effective and safe way. The 3D RUS-PAT system consists of four main parts, as shown in Fig. 1: i) a single-element ultrasonic transducer with pulser for transmitting ultrasonic waves, ii) a laser to direct light to the object, iii) arc-shaped ultrasonic transducer arrays evenly distributed on a hemispherical bowl with an azimuthal scanning mechanism to record scattered ultrasound and photoacoustic signals panoramically, and iv) a signal amplification, data acquisition (DAQ), and processing system which amplifies and digitizes the signal and then reconstructs a volumetric image. The positioning of the human subjects during various imaging sessions (head, breast, hand, and foot) is described in the Methods section and illustrated in Supplementary Fig. 1.

In the RUST implementation, we used a water-immersed 2.25 MHz single-element ultrasonic transducer as the ultrasound source. It was driven by an ultrasound pulser which consists of a function waveform generator (RIGOL, DG1022, 20 MHz, 100 MSa/s) and a power amplifier (ENI 240 L RF Power Amplifier, 20 kHz to 10 MHz, 50 dB) to generate a wide-field spherical wave from focus to the object (blue color in Fig. 1). In one of the example implementations, the detection array has four 256-element quarter-ring arrays distributed on a hemispherical housing surface with a pitch of 90 degrees (center frequency 2.25 MHz, Imasonic Inc.) which is azimuthally rotated using a stepper motor for panoramic detection. The number of arc-shaped arrays is flexible and can be adjusted based on the trade-off between imaging speed, scanning angular range, and system cost. The source transducer was mounted co-axially with the stepper motor at the intersection of the arc-shaped detection arrays. The co-axial azimuth scanning mechanism ensures that the location of the ultrasound source does not change with the rotation of the detection arrays. With 90 degrees of continuous rotational scanning, we achieve panoramic synthetic hemispherical ultrasonic detection, which is equivalent to a physical hemispherical detection array in a cost-effective way. The detected signals are amplified and streamed to the computer through the DAQs, and they form RUST images of the human body.

The RUST system is integrated with a 3D PAT implementation[21] to achieve a hybrid RUS-PAT system. When switched to the PAT mode, an engineered diffuser within a lens tube is installed at the intersection of the arc-shaped detection arrays instead of the source transducer in RUST. We use a 1064 nm laser with an engineered diffuser to direct a diffused light beam to the object and generate photoacoustic signals. PAT shares the same arc-shaped detection arrays, a rotational motor, DAQs, and a processing system with RUST. The hybrid 3D RUS-PAT system is operated using custom software that enables dual working modes – RUST and PAT. The acquisition can start with the PAT mode, followed by the RUST mode, to obtain a dual-modality image of the human body. The data acquisition and mode switching methodologies are described in Methods and in Supplementary Fig. 2. Additionally, we also implemented a RUS-PAT version that can perform simultaneous RUST and PAT, as shown in Supplementary Fig. 14. In this setup, we deliver the light from the side of the array and install a single ultrasound transducer at the bottom of the arc-shaped array. Thus, we do not need to switch the ultrasound transmission transducer and the diffuser of the light but can obtain dual image modalities of the targets without changing the imaging position. It shows comparable image features of a light-absorbing and ultrasound-scattering phantom images by PAT with bottom illumination, PAT with side illumination, and RUST, respectively. The slight discrepancy between the phantom images obtained from bottom and side illumination is due to the use of different lasers, with a 20 Hz laser (Litro, PRF: 20 Hz, maximum pulse energy: ~2.5 J) used for bottom illumination and a 10 Hz laser (Quantel, PRF: 10 Hz, maximum pulse energy: ~850 mJ) for side illumination.

**Phantom validation of RUS-PAT.** We performed phantom experiments to evaluate the performance of the 3D RUS-PAT system. One example of the array setup for RUST is shown in Fig. 2a. The single-element source transducer is at the center (arrow) of the detection array which has four arc-shaped arrays. The system achieved a FOV of ~10 cm and an isotropic spatial resolution of ~ 400 µm (along $x$, $y$, and $z$ directions) for both RUST and PAT as shown in Fig. 2b. The field of view (FOV) is determined by the region of the source acoustic field and the light illumination, which is smaller than the diameter of the arc-shaped arrays. The total scanning time to acquire each volumetric image is 10 s. The images of a metal wire cross phantom at a depth of 2 cm obtained using RUST and PAT are shown in Figs. 2c and 2d respectively. We were able to observe comparable structures and the width of the object (~ 1.5 mm indicated by the profiles in Fig. 2c-d) with RUST and PAT. A comprehensive assessment of the spatial resolution uniformity for RUST and PAT across the image FOV is presented in Supplementary Fig. 3 and 4, respectively. In the case of RUST, the maximum percentage changes in spatial resolution along the $x$, $y$, and $z$ directions are found to be 7.8%, 5.1%, and 3.5%, respectively. As for PAT, the corresponding maximum percentage changes in spatial resolution along the $x$, $y$, and $z$ directions are 6.4%, 5.8%, and 6.3%, respectively.

To quantitatively evaluate the sensitivity distribution within the image FOV, we conducted a simulation study to assess the signal-to-noise (SNR) change, considering acoustic signal attenuation in soft tissue. The results are shown in Supplementary Fig. 5 (RUST) and Supplementary Fig. 6 (PAT). The maximal percentage change of SNR was 55% for RUST and 58% for PAT from an acoustic perspective, respectively. However, for PAT when we considered both light and acoustic attenuation in soft tissue, the distribution of SNR in the FOV shows a drop of approximately 50 dB with a 4 cm penetration depth. Additionally, we also imaged a light-absorbing and ultrasound-scattering target to experimentally evaluate the system's sensitivity and spatial resolution distribution, as shown in Supplementary Fig. 7.

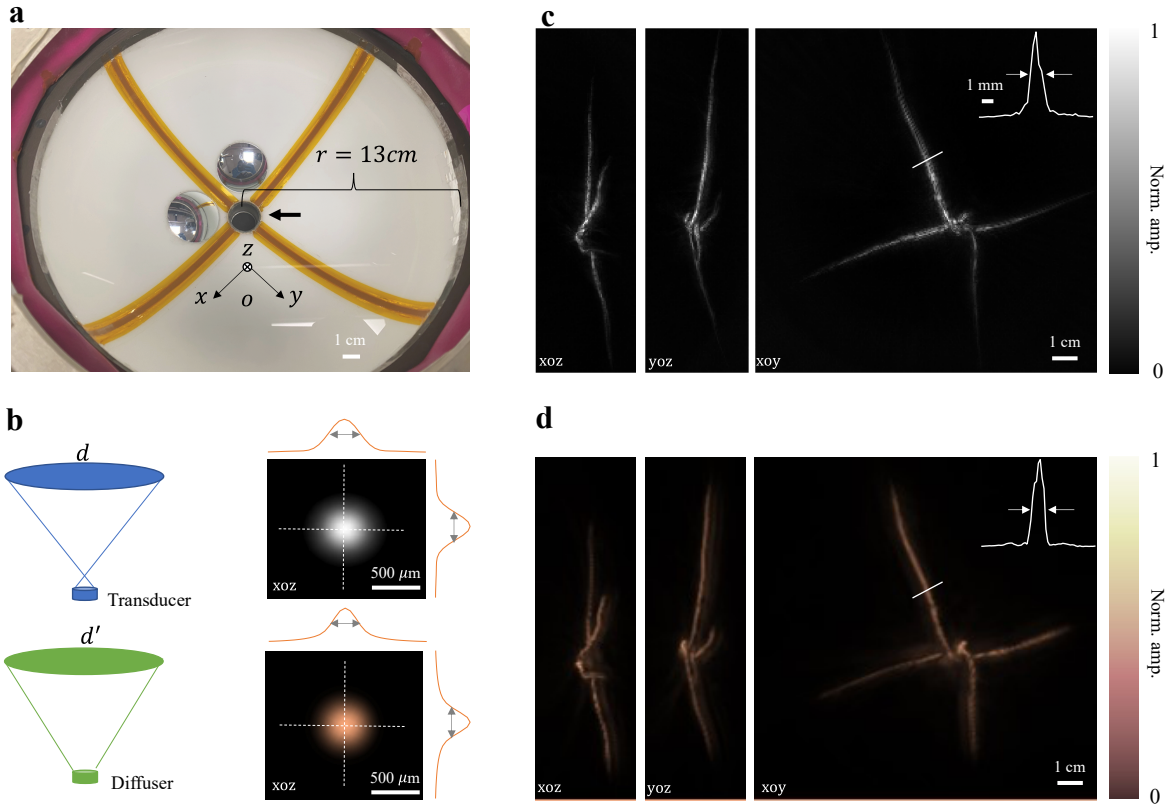

**Fig. 2 | Array and phantom images**. **a**, An example RUST setup with four 256-element quarter-ring arrays and a single element transducer (arrow). Note that we added two silver mirrors to help to adjust the position of the imaging target. **B**, Field of view for RUST ($d \approx 10$ cm) and PAT ($d' \approx 10$ cm) setup, and isotropic spatial resolution around 400 μm (along the $x$, $y$, and $z$ directions) for RUST (top) and PAT (bottom). **c-d**, a light-absorbing and ultrasound-scattering phantom was imaged using the 3D RUS-PAT system in (**c**) RUST mode and in (**d**) PAT mode. The images are 12.5 cm × 12.5 cm × 3.0 cm.

**RUS-PAT of the human head *in vivo*.** Imaging the morphological and angiographic information of the head is valuable for the evaluation of head injuries[28,29], diagnosis of brain disease[30,31], and the study of brain function[14]. We imaged the skull-less hemisphere of a hemicraniectomy patient's head with RUS-PAT. The maximal amplitude projection (MAP) images of the head by 3D RUST (Fig. 3a) reveal the scalp boundary and muscular structures in the head. Images from more views of the head are also shown in Supplementary Fig. 8, and they show a clear boundary between the scalp and the cortical region of the brain (white arrows in Supplementary Fig. 8). Brain angiography images were obtained in the PAT mode (Fig. 3b), which shows the vascular network in the head. The PAT signals are related to brain function because PAT directly measures the content of hemoglobin. In addition, the patient has a surgical suture area in the head. Due to the light-absorbing and ultrasound-scattering contrasts, we could resolve the suture area by both RUST and PAT (white arrows), and they co-registered well with each other. Overall, the 3D dual-contrast head images presented here reveal rich information for potential evaluation of head injuries, diagnosis of brain diseases, and the study of brain function.

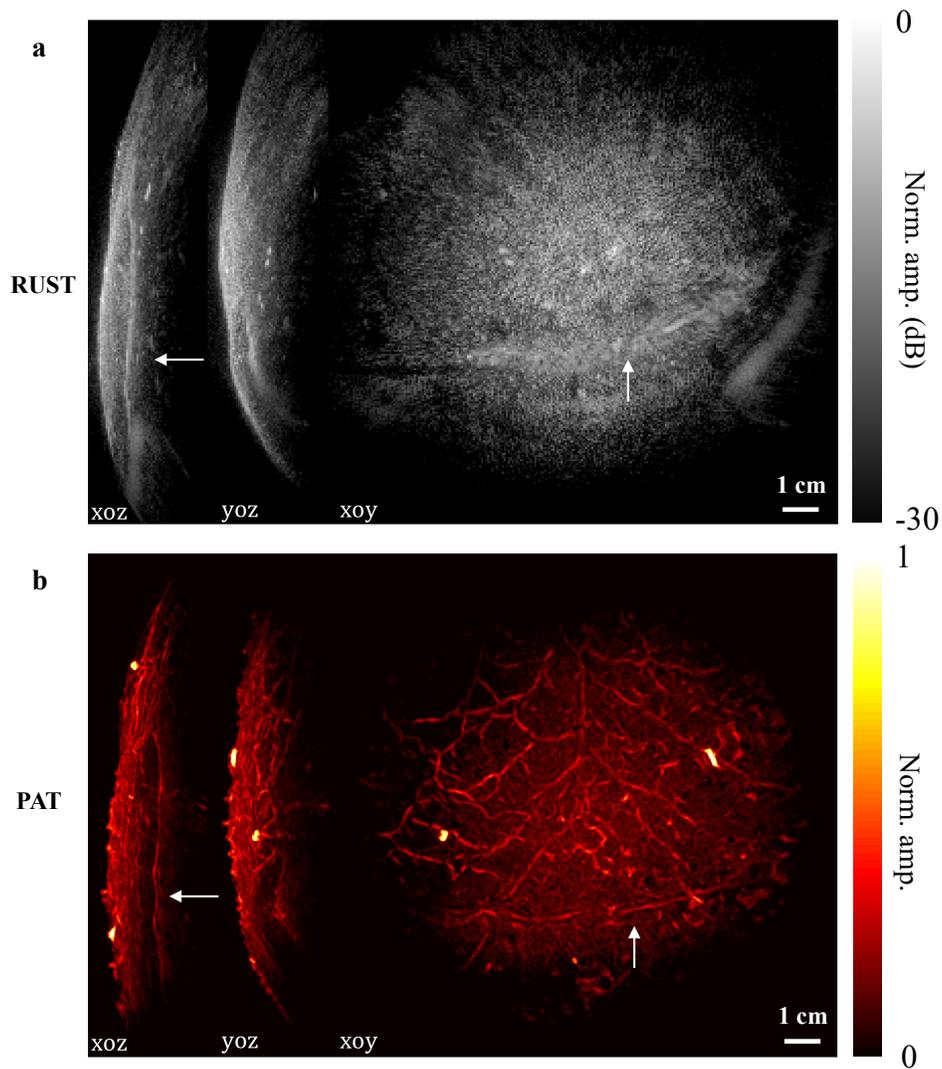

**Fig. 3 | RUS-PAT images of the head. a**, MAP images of the head by RUST. **b**, MAP images of the head by PAT. The images are 12.5 cm × 12.5 cm × 3.0 cm. Bright spots are the markers for PAT. The bright points in the PAT image refer to the markers on the surface of the head to assist with the co-registration of the photoacoustic images with the photograph of the head's region of interest. White arrows indicate the surgical suture area in both the RUST and PAT images.

**RUS-PAT of the breast *in vivo*.** Imaging the whole-breast soft tissue and vasculature is a medical need for the monitoring and diagnosis of breast cancer[33,34]. We imaged the breast of healthy subjects in a prone position using the 3D RUS-PAT system. We started with the PAT mode and acquired data within a single breath-hold of 10 seconds. Next, with the subject in the same position, we switched the system to the RUST mode and acquired RUST data for another 10 seconds. We demonstrated that RUS-PAT could provide soft tissue ultrasound structural images (Figs. 4a-b, d-e) and PAT angiographic images (Figs. 4c, f) quasi-simultaneously. In the RUST images (Figs. 4a-b, d-e), we visualized the 3D ultrasound structures of the left and right breasts, from the skin (indicated by white arrows) to deep internal features (indicated by orange arrows). Additional MAP images and slice views of more subjects are shown in Supplementary Fig. 9c-k, providing a clear visualization of the internal structure of breast tissue from the nipple to the chest wall. In the PAT images (Figs. 4c, f), blood vessels inside the breast are well resolved. Large blood vessels are located near the skin (indicated by white arrows), and smaller blood vessels are deeper in the breast (indicated by orange arrows), which is consistent with the general angiographic anatomy of the breast. Zoomed-in side-by-side comparisons of the RUST and PAT images are shown in Supplementary Figs. 9a-b. The dual-modality RUS-PAT images have the potential to measure the volume of the tumor using RUST and visualize the vasculature inside and close to the tumor through PAT in breast cancer patients[35].

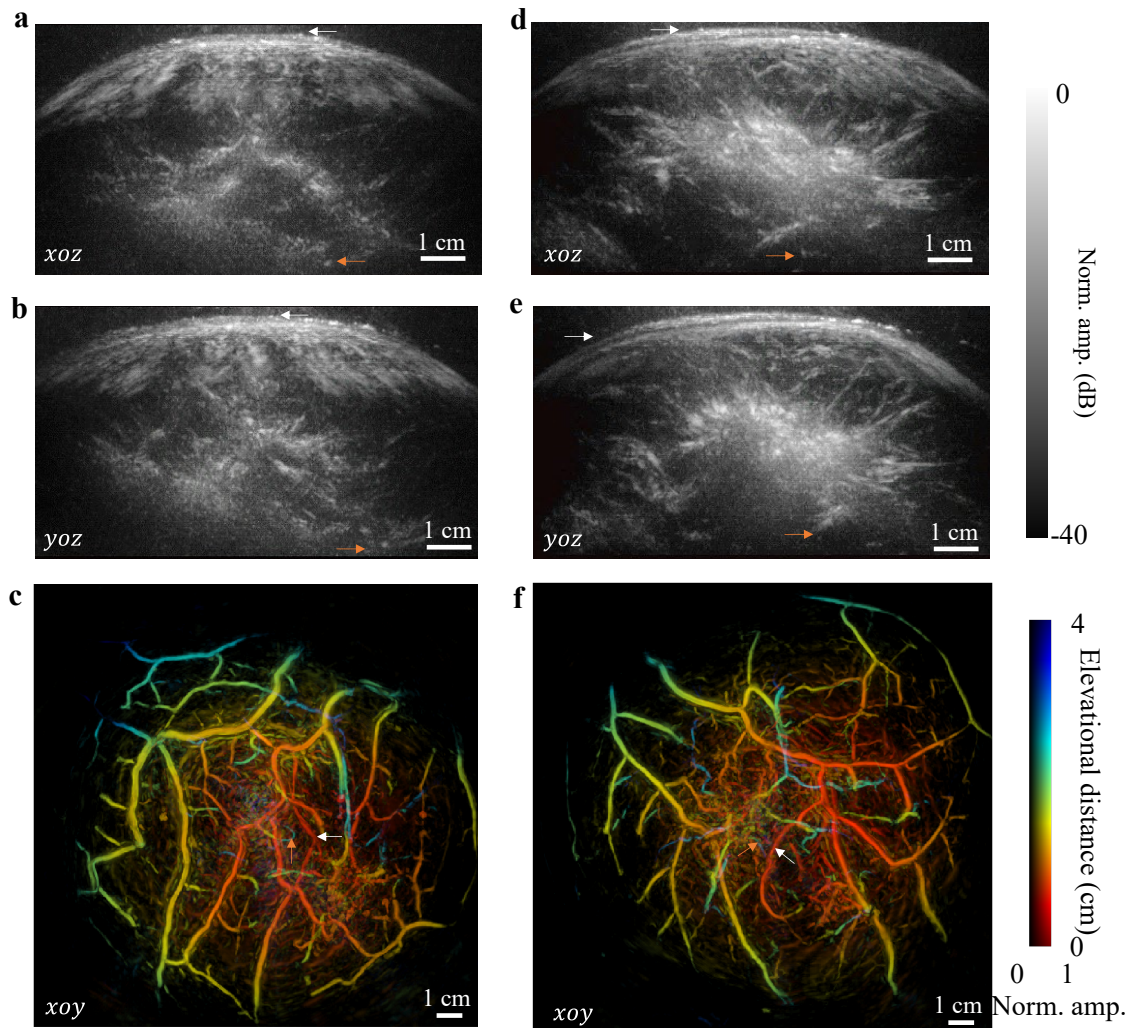

**Fig. 4 | RUS-PAT images of the breast of a healthy female subject. a-b**, MAP images of the left breast by RUST. **c**, Elevationally-encoded MAP image of the left breast by PAT. **d-e**, MAP images of the right breast by RUST. **f**, Elevationally-encoded MAP image of the right breast by PAT. White and orange arrows in the RUST images (**a-b, d-e**) indicate the skin surface and deep internal structures at a depth of approximately 5.0 cm from the skin surface, respectively. White and orange arrows in the PAT images (**c, f**) indicate superficial vessels close to the skin surface and deep vessels at a depth up to approximately 3.2 cm from the skin surface, respectively.

**RUS-PAT of the extremities *in vivo*.** Imaging the muscular tissue and vasculature of the extremities helps in evaluating injuries and screening for peripheral neurovascular diseases[36,37]. We imaged the hand palm of a healthy subject and showed the soft-tissue structure captured by RUST (Fig. 5a) and the major and minor vasculature captured by PAT (Fig. 5b), respectively. In Fig. 5b, the vessel network of the hand can be clearly tracked inside the muscular tissues. More RUS-PAT images for the whole palm, fingers, and back of the hand are shown in Supplementary Fig. 10. Additionally, we imaged the dorsal side of both the left and right foot of a subject. The right foot had an open wound on its dorsal side. Figs. 5c and 5d depict the soft-tissue structure captured by RUST and the vasculature captured by PAT of the healthy left foot, respectively. The white arrows in Fig. 5c indicate the metatarsal regions of the foot. In Fig. 5e, a bright region indicated by the dashed white circle represents the wound on the dorsal side of the right foot in the RUST image. Similarly, Fig. 5f shows the wound region in the PAT image, also indicated by a white circle. Our results indicate that 3D RUS-PAT has the potential to detect tissue abnormalities by RUST and vascular abnormalities by PAT in extremities with injuries or disease.

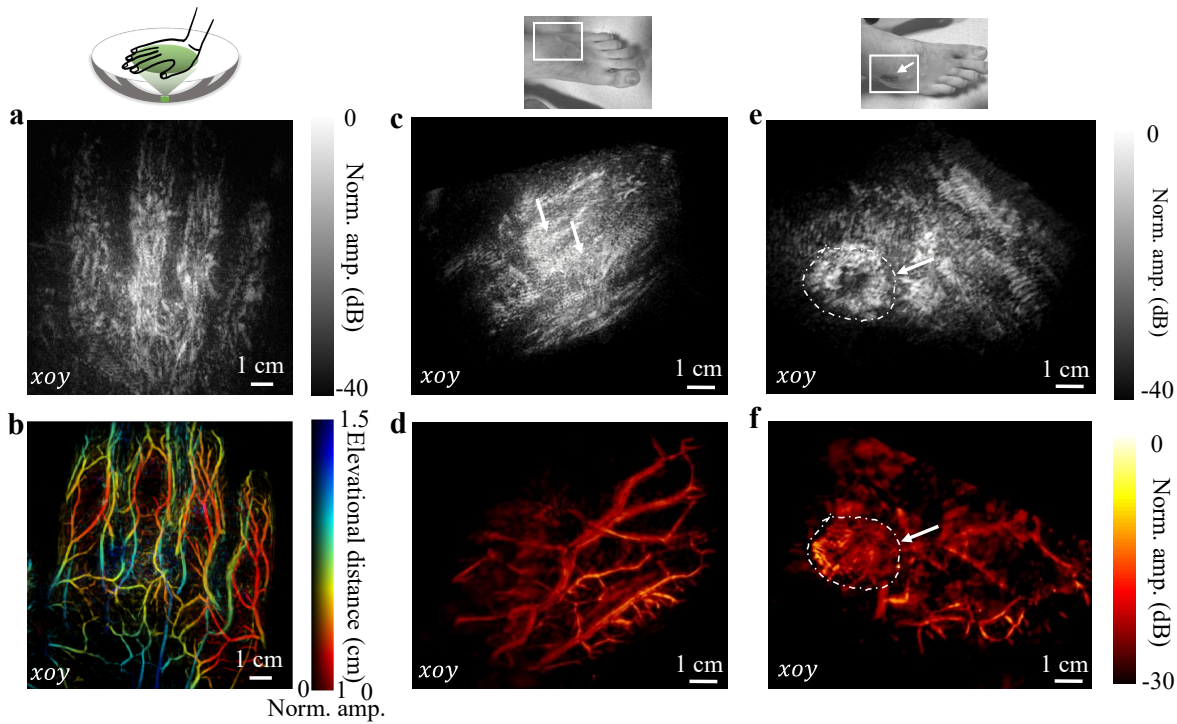

**Fig. 5 | RUS-PAT images of the hand and foot. a**, MAP images of the hand by RUST. **b**, MAP images of the hand by PAT. The images are 10.0 × 10.0 × 1.5 cm. **c**, MAP images of the dorsal side of the left foot by RUST. The white arrows indicate the metatarsal regions of the foot. **d**, MAP images of the dorsal side of the left foot by PAT. **e**, MAP images of the dorsal side of the injured right foot by RUST. **f**, MAP images of the dorsal side of the injured right foot by PAT. The white arrows and dashed white outlines (**e, f**) represent the wound on the foot. The white boxes in the photographs represent the respective regions of the feet being imaged.

**RUS-PAT of a patient with diabetic foot *in vivo*.** Diabetic foot ulcers are open wounds that develop on the feet of people with diabetes due to nerve damage, poor circulation, and impaired healing. Without proper care, they can lead to infections and even amputation. We imaged a patient with diabetic foot ulcers using our RUS-PAT system. Figs. 6a-c depict the photograph, the soft-tissue structure captured by RUST, and the vasculature captured by PAT of the normal left foot, respectively. The right foot of the patient had an ulcer on the bottom and only four toes. In Figs. 6e and h, a bright region indicated by the square box and dashed white circle represents the foot ulcer on the bottom of the right foot in the RUST image. Similarly, Figs. 6f and i show the foot ulcer region in the PAT image, also indicated by the square box and a white circle. Additionally, we observed the loss of the thumb toe's feature in RUST (Fig. 6e) and PAT (Fig. 6f), which matches the photograph (Fig. 6d). Note that the bright spots, as indicated by the solid white arrows in Fig. 6f, correspond to the surgical suture region of the foot, and the features, as indicated by the dashed white arrows in Fig. 6e, correspond to the band tape on the surface of the toes. Our results indicate that 3D RUS-PAT has the potential to detect diabetic foot ulcers in patients.

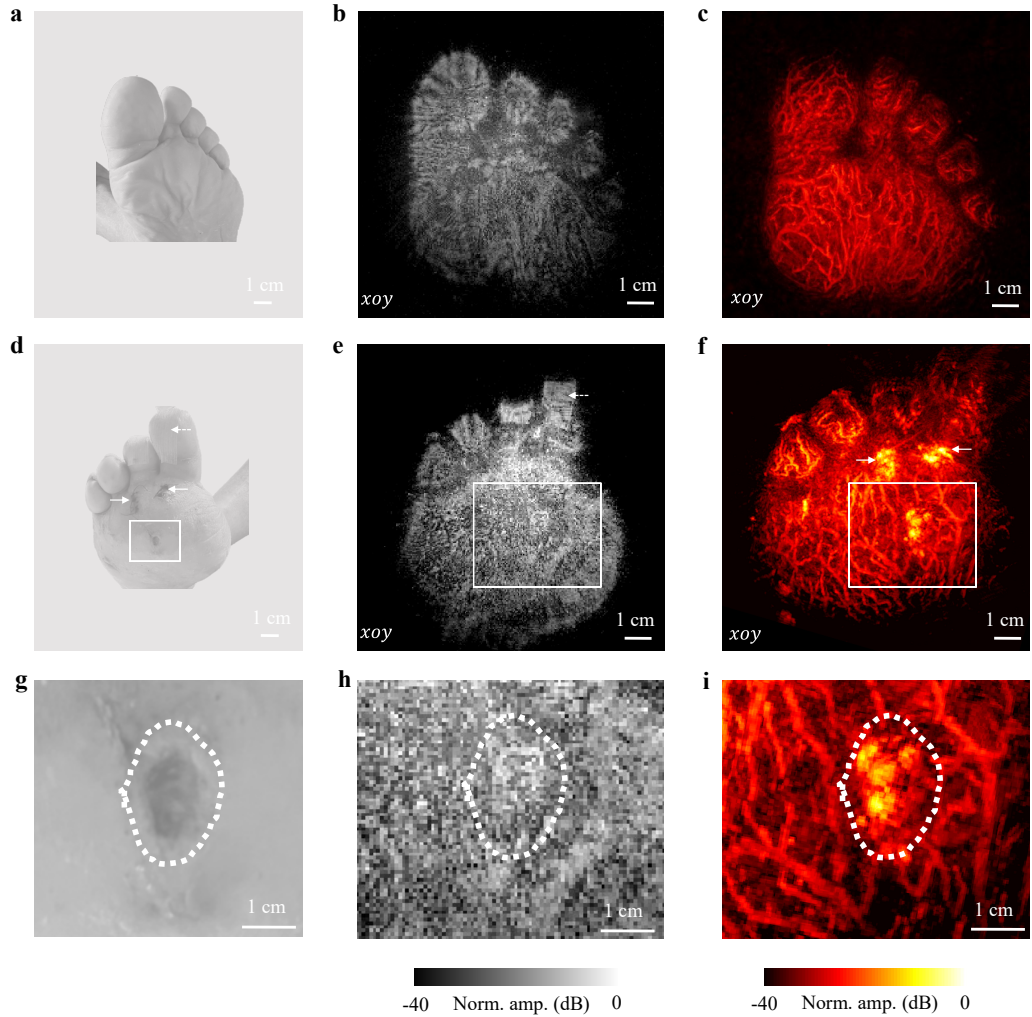

**Fig. 6 | RUS-PAT images of a patient with diabetic foot ulcers. a-c,** Photograph, MAP RUST images, and MAP PAT images of the normal left foot, respectively. **d-f,** Photograph, MAP RUST images, and MAP PAT images of the abnormal right foot, respectively. **g-i,** Zoomed-in photograph, MAP RUST images, and MAP PAT images of the abnormal right foot, respectively.

## Discussion

By imaging both morphological and angiographic information quasi-simultaneously with a large FOV (~ 10-cm diameter), submillimeter isotropic resolution (~ 400 μm), and high-speed (10 s), RUS-PAT has the potential for rapid clinical translation. A major advantage of RUS-PAT for human medical imaging is that it requires neither exogenous contrast agents nor ionizing radiation, improving safety for the patient. In this study, we achieved volumetric ultrasound structural and PAT angiographic images in the human head, breast, hand, and foot. For the head imaging, we imaged a hemicraniectomy patient's skull-less hemisphere because the skull presents a major obstacle for transcranial ultrasound tomography and PAT with a 2.25 MHz ultrasound center frequency of the current system. Here, the intention was to showcase the capabilities of the system in imaging brain tissue and detecting potential abnormalities or pathologies. A lower ultrasound center frequency can be utilized for transcranial application in the future. For the breast application, we imaged healthy breasts to demonstrate the capabilities of RUS-PAT to obtain the soft tissue structures and blood vessels inside the whole breast. Evaluation of the performance of RUS-PAT in a breast cancer patient would be of major interest. One of the expectations is that RUST can detect the location and measure the volume of the tumor, and PAT can obtain the vasculature distribution of the tumor region, which will benefit the early diagnosis of breast cancer[10]. For the hand images, the morphological and angiographic information of the hand can be potentially used for biometric identification in healthy subjects[38,39]. RUS-PAT also has the potential to be used for the evaluation of hand injuries, to detect structural pathology such as a cyst in the hand, to monitor postoperative limb reconstruction/salvage outcomes, and for the evaluation of peripheral vascular disease[40]. Another impactful application in the extremities might be with diabetic foot, where chronic ischemia, wound tissues, and infections make RUS-PAT a potential clinical imaging technique. We imaged a patient with diabetic foot ulcers on the bottom of the foot. Our results demonstrate that RUS-PAT can detect tissue abnormalities through RUST and vascular abnormalities through PAT in patients with diabetic foot ulcers. In addition, RUS-PAT can also be used for other parts of the human body, such as the abdomen, by adjusting the position of the system and the subject. In addition, free flaps in plastic surgery might be another application where RUS-PAT might find immediate value in the critical first few days after surgery.

Several key features make RUS-PAT a unique tool for 3D dual-contrast imaging. The rotating mechanism ensures that we can use arc-shaped arrays to synthesize a hemispherical array with a panoramic detection view for both RUST and PAT. The use of a separated single-element ultrasonic transducer for transmission in RUST ensures that it is highly compatible with the PAT system, which has a detection-only front-end circuit. The use of a single-element transmission transducer and arc-shaped detection arrays makes RUS-PAT achieve panoramic 3D imaging in a cost-effective way because it does not require a physical hemispherical array with a large number of array elements and their corresponding detection channels. The PAT mode uses the same detection module as the RUST mode, which reduces system complexity. We have developed a switch methodology (Supplementary Fig. 2) to ensure the subject remains in the same position during the RUST and PAT modes. The system is also capable of performing RUST and PAT simultaneously by delivering light from the side of the array and installing a single ultrasound transducer at the bottom of the arc-shaped array (Supplementary Fig. 14).

The 3D RUS-PAT system is flexible with the choice of the ultrasound center frequency, the number of arc-shaped arrays, and the imaging time. In the current setup, we used a detection array with a center frequency of 2.25 MHz. It is possible to use a much higher transducer frequency, such as 5 MHz, for an improved spatial resolution for human breast, hand, and foot imaging at the cost of reduced penetration depth. Similarly, we can use a 1 MHz center frequency for transcranial human head imaging at the cost of reduced spatial resolution. The choice of the number of arc-shaped arrays depends on the trade-off between imaging speed, signal-to-noise ratio, and hardware cost. We can reduce the number of arcs by using a single-arc setup but with a 360 degrees mechanical scanning angle. This setup requires four times the scanning time but reduces the number of array elements and DAQ channels by four times compared with a four-arc setup, thus reducing the cost further. We can also increase the number of arc-shaped arrays to 8, 16, or even more to increase the imaging speed to capture functional information and image fast phenomena in the human body with increased hardware cost. The scanning time can be reduced to 2 s when we operate a higher speed motor and use a higher pulse repetition rate (i.e., 50 Hz) for ultrasound and light transmission to ensure a densely sampled hemispherical detection matrix. With such a high speed, single breath-hold imaging of different parts of the human body can be achieved with minimal motion during acquisition. Furthermore, in the current implementation, the ultrasound and light source have been installed at the bottom of the array. However, it is worth noting that we have the flexibility to adjust their positions to other locations, such as the side of the array. This alternative configuration holds the potential to serve as additional sources, thereby enhancing the SNR within the entire volume.

In this study, we have successfully developed a proof-of-concept system for imaging representative areas of the human body. Our system enables the acquisition of both morphological and angiographic information in a quasi-simultaneous manner, offering a large field of view (approximately 10-cm diameter), submillimeter isotropic resolution (around 400 µm), and high-speed imaging capabilities (10 seconds) for regions such as the human head, breast, hand, and foot. However, it is important to acknowledge the limitations and the direction of further improvement of our current study.

1) The sensitivity for detecting small blood vessels. We performed the analysis of the sensitivity for detecting small blood vessels in the head, breast, and extremities, as shown in Supplementary Fig. 11. The measured smallest diameters of the blood vessels of the hand, breast, and head in the PAT images are $0.62 \pm 0.06$ mm (at a depth of 1.0 mm), $0.60 \pm 0.03$ mm (at a depth of 1.0 cm), and $0.61 \pm 0.10$ mm (at a depth of 3.0 mm), respectively. Note that the *in vivo* blood vessels may be smaller than the measured value due to the spatial resolution of the system. While our current system demonstrates the capability to detect submillimeter blood vessels, there is room for further improvement. To enhance the spatial resolution, we could explore the use of a wider ultrasound band, which would enable us to delineate even smaller blood vessels below one hundred micrometers in diameter. Additionally, employing SNR boosting techniques such as incorporating additional sources could contribute to better imaging quality and more precise visualization of these tiny blood vessels. Thus, by employing these techniques we may achieve even greater resolution and sensitivity in our imaging system.

2) Penetration depth and volumetric coverage. In our current implementation, we have achieved a penetration depth of 5 cm with RUST imaging, as demonstrated in Fig. 4 and Supplementary Fig. 9. For breast imaging using PAT, we have achieved a penetration depth of 3.2 cm in one female subject, as demonstrated in Fig. 4. The slightly shallower penetration depth of the breast compared to a previous study[22] is due to the use of water instead of deuterium oxide ($D_2O$). These depths allow us to access the cortex region of the human brain, a significant portion of the breast, and the entire hand. However, there is potential for further exploration to observe deeper features, particularly in deep brain imaging and whole breast imaging to the chest wall. By pushing the depth limit (e.g., the use of lower center frequency transducer arrays and optimal laser wavelengths), we may capture more comprehensive information from these regions. Furthermore, our system offers a field-of-view with a diameter of 10 cm in the horizontal direction. This field-of-view is sufficient for capturing images of most parts of the head, breast, hand, and foot. Note that this field-of-view can also be adjusted by changing the diverging angle of the engineering diffuser and the position of the source ultrasound transducer. Moreover, to enable whole-body imaging, particularly for the abdomen region, it is possible to enlarge the field-of-view by utilizing a larger diameter for the arc-shaped array. This enhancement would facilitate more comprehensive imaging coverage of the entire body.

3) System adaption for different target sizes. The transducer array used in our RUS-PAT system was designed with a 13 cm radius. This size was selected to accommodate a wide range of head sizes and breast sizes typically encountered in clinical practice. The 13 cm radius offers sufficient coverage for most individuals, allowing adequate imaging of the target area of interest. We added Supplementary Figs. 1 and 12 to illustrate the experimental setup, in which the target was supported by the plastic film, which can support various sizes and geometries. While the fixed size of the transducer array provides versatility in accommodating various head and breast sizes, we acknowledge that there may be limitations for extremely large or small anatomical structures. Future work includes the design of more advanced holders for different sizes of head, breast, and other body regions.

4) Comparison of RUST with clinical handheld ultrasound. The large three-dimensional field-of-view (10 cm diameter) offered by RUST enables whole organ imaging, such as the entire breast or hand, with a single acquisition. In contrast, conventional handheld linear array probes typically provide only two-dimensional images with limited field-of-view (less than 5 cm). The handheld operation of conventional probes allows clinicians to explore the region of interest through multiple trials. However, this flexibility can introduce operator-dependent imaging results and discrepancies between visits due to variations in operations and imaging locations. In contrast, our RUST system overcomes these limitations by capturing whole breast and hand images in a single acquisition, potentially providing more reliable and consistent results compared to handheld operations. While the current implementation of RUST is not designed for handheld use, we acknowledge the value of a handheld version. In fact, we envision the possibility of implementing the RUST concept in a handheld probe by utilizing a smaller diameter arc array and a lightweight rotational motor, accompanied by a longer electrical cable for connection to the data acquisition system. This handheld version could serve as a complementary bedside application, providing clinicians with the flexibility they are accustomed to, albeit with a relatively smaller field-of-view. We understand that adopting new technology in a clinical setting can present challenges. However, we firmly believe that with proper training and familiarization, clinicians can effectively adapt to using RUST. The operation of RUST is similar to other existing medical imaging devices, such as CT, MRI, and ultrasound systems. Therefore, clinicians can leverage their existing expertise and familiarity with these modalities to easily integrate RUST into their clinical workflow.

For the ultrasound mode of RUS-PAT, we have included results demonstrating breast imaging using our current 2.25 MHz system in comparison with a standard linear array probe (ATL P4-2, center frequency 2.5 MHz), with both operating approximately at the same center frequency. We selected the same center frequency for the linear array probe to ensure a fair comparison, as the RUS-PAT method is adaptable to different center frequencies. As presented in the added Supplementary Fig. 15, a standard linear array probe was used to image the same breast subject, matching the center frequency of the RUS-PAT system. Two comparable planes were selected from the 2D slices of the three-dimensional RUST images and the corresponding linear array images for comparison. The linear array imaging sequence was performed using standard B-mode focused ultrasound imaging. By comparing Supplementary Fig. 15b and c, as well as Supplementary Fig. 15e and f, it is evident that RUST provides significantly better spatial resolution and more precise delineation of breast structures than the standard linear array probe. The slight discrepancies between Supplementary Fig. 15b and c, as well as Supplementary Fig. 15e and f, can be attributed to gentle compression of the breast by the linear array probe and positional differences during imaging.

We have also added a comparison of RUS-PAT in photoacoustic mode with state-of-the-art Doppler ultrasound imaging. As shown in Supplementary Fig. 16, we acquired hand images using RUS-PAT, ultrafast Doppler ultrasound with a matrix array probe (1024 elements, 0.3 mm pitch, 8 MHz center frequency, Vermon, Verasonics, Inc.), and ultrafast Doppler ultrasound with a linear array probe (256 elements, 15 MHz center frequency, LZ250, VisualSonics Inc.). One significant advantage of RUS-PAT is its large field-of-view, approximately 10 cm in diameter (Supplementary Fig. 16a), compared to the matrix array's ~1 cm (Supplementary Fig. 16b) and the linear array's ~2 cm 2D slice (Supplementary Fig. 16c). The regions imaged by the matrix array and linear array Doppler ultrasound correspond to the white box and dashed green line in Supplementary Fig. 16a, respectively. The images clearly demonstrate that RUS-PAT captures the majority of the hand region and provides a detailed view of the blood vessel network. In contrast, Doppler ultrasound captures only two blood vessels with the matrix array or a single blood vessel with the linear array. Another key difference between RUS-PAT and state-of-the-art Doppler ultrasound lies in the physical meaning of the images. RUS-PAT measures relative hemoglobin content, whereas Doppler ultrasound detects blood flow. We believe RUS-PAT can serve as a highly complementary tool to Doppler ultrasound, providing rich and complementary information about the circulatory system.

5) Image quality and artifacts. In the case of breast imaging, the dark gap observed in the image (Fig. 4a) is due to the relatively weak scattered ultrasound signals from the corresponding breast tissue. In the case of extremity imaging, we observed striation patterns on the surface of the foot in the RUST images (Fig. 6b). These patterns can be attributed to the geometric characteristics of the tissue surface (as indicated by the photograph Fig. 6a) and may also be influenced by speckle artifacts. To address this, further exploration of data processing techniques is warranted (e.g., speckle reduction). These techniques could potentially help in removing or utilizing this pattern to extract additional information from the extremity images. Additionally, we chose to use a 30 dB or 40 dB log compression threshold for displaying the ultrasound images, as it provides a good balance between image contrast and detail in our case. Achieving higher image contrast or dynamic range (up to 60 dB) could be possible with advanced hardware configurations (e.g., higher center frequency ultrasound arrays) or more sophisticated data processing techniques (e.g., adaptive image reconstruction methods).

6) Overall cost. To analyze the overall cost of the current RUS-PAT system implementation, we need to consider the arc-shaped arrays, the pre-amplifiers, the data acquisition modules, the laser, and the host computer. One of the significant contributions of our work is that we have offered a cost-effective solution for adding ultrasound imaging modality to the photoacoustic computed tomography system. In detail, the rotational ultrasound tomography concept is highly compatible with photoacoustic tomography because it uses a separate single-element transducer for transmission and, therefore, does not require significant hardware modification. In contrast, adding conventional ultrasonography to a photoacoustic system requires the addition of hundreds or thousands of pulsers and switches due to photoacoustic tomography's detection-only front-end circuit. Additionally, we have exploited the rotational mechanism with arc-shaped detection arrays to achieve three-dimensional panoramic ultrasonic detection for both ultrasound and photoacoustic tomography in a cost-effective way from the ultrasound array perspective, instead of using a high-cost, dense, hemispherical array. Since our current implementation does not introduce a new method for reducing the cost of the PAT component, we acknowledge that the overall system cost is largely determined by this component, which is a limitation of current PAT systems. To address this issue, future efforts may focus on leveraging low-cost light delivery (e.g., LED array) and acoustic detection array technologies (e.g., a single arc-shaped array, at the cost of increased imaging time).

7) Potential for future clinical application. The current study introduces an accessible and adaptable implementation of RUS-PAT that integrates 3D ultrasound tomography with photoacoustic tomography. While the system is in an early stage, its modular design offers flexibility for future optimization toward different clinical needs. In breast imaging, the current RUS-PAT setup has enabled dual-modality 3D ultrasound and photoacoustic angiography images in healthy volunteers, with depth coverage sufficient to capture a substantial portion of the breast, including areas of potential clinical interest. With further system refinements—such as using higher-frequency ultrasound arrays and improved illumination strategies—there is potential to enhance image resolution and coverage for more comprehensive breast assessment. In detail, with additional modifications—such as breast compression and dual-sided illumination—full-breast PAT may become feasible. Additionally, it is important to note that current clinical breast ultrasound often utilizes higher-frequency probes (8–15 MHz), whereas our prototype uses a 2.25 MHz array. Future versions may incorporate higher-frequency arrays to enhance spatial resolution, although this may limit imaging depth. In head imaging, the ability to detect cranial sutures demonstrates sensitivity to anatomical landmarks, suggesting feasibility for exploring superficial cerebral vasculature in future studies. For hand imaging, the system shows promise in visualizing structures such as blood vessels, tendons, and joints, which may support future investigations into musculoskeletal and vascular conditions. Additionally, initial application of the system in patients with diabetic foot ulcers indicates potential for assessing tissue damage and vascular changes. While these early demonstrations are encouraging, further validation with larger cohorts and condition-specific adaptations will be necessary to fully evaluate the system's clinical relevance across these applications.

In this pilot study, we demonstrated RUST's ability to achieve a large FOV anatomical image at high speed and its flexibility to be compatible with PAT as RUS-PAT for obtaining both the ultrasound structural and PAT angiographic images of the human body. This state-of-the-art dual-contrast RUS-PAT system provides volumetric imaging of the human body in a high-speed, large 3D FOV, cost-effective and safe way, which is viable for clinical translation.

## Methods

**3D RUST design and construction.** The 3D RUST design comprises a source ultrasound transducer, an ultrasound detection module, and a rotation scanner. The ultrasound source was a spherical focused single element transducer (Olympus, focal length = 1 inch, diameter = 0.75 inches, 2.25 MHz center frequency). The transducer generates a virtual point source in front of the transducer surface. In one of the example demonstrations, ultrasound detection was implemented by four 256-element quarter-ring ultrasonic transducer arrays evenly distributed on a hemispherical bowl (130-mm radius) with a separation of 90 degrees. The array elements are unfocused with a 0.6 mm × 0.7 mm active element size, 0.74 mm element pitch, and 2.25 MHz center frequency with a one-way 78% fractional bandwidth. The number of arc-shaped arrays used is flexible and can be adjusted to 1, 2, 4, 8, or even more. The 1024 detection elements are directly connected to four 256-channel lab-made preamplifier modules (gain settings of 21 dB). The amplified acoustic signals are acquired by four 256-channel DAQ boards (Photosound Inc.) with a 20 MHz sampling rate, gain settings of 20 dB, and streamed to a workstation by USB 3.0 in real-time. A physical picture of the system and its main components are shown in Supplementary Fig. 12. The ultrasound transducer array, motor, pre-amplifiers, and DAQs are conveniently installed beneath the bed, allowing the subject to comfortably lie down during system operation. The arc-shaped arrays with a hemispherical housing are positioned below the bed to ensure effective coupling with the imaging targets. Four 256-channel DAQ boards (Photosound, Inc.) are integrated on the side of the rotational motor. These boards capture and amplify the acoustic signals, enabling further signal processing.

The source ultrasound transducer was positioned at the intersection of the arc-shaped detection arrays which are mounted on a co-axial stepper motor (NEMA 34, 8V) coupled with a set of two spur gears (Designations, Inc., KSS2-20J12 and KSS2-120, gear ratio = 1:6). The stepper motor scans in the azimuthal direction to achieve a 3D panoramic ultrasound detection. The co-axial design of the source ultrasound transducer, arc-shaped detection arrays, and the scanning motor ensures that the virtual point source remains in the same location when the motor scans in the azimuthal direction. In this case, we can assume a consistent acoustic field distribution in the FOV while scanning the ultrasonic arrays for the image reconstruction of 3D RUST. We chose to use a single-element source transducer to reduce the system cost and complexity compared with the use of a large number of pulses and transmission array elements. The arc-shaped detection arrays design with a rotation scanner ensures sufficiently dense elevational and azimuthal sampling to generate a synthetic hemispherical array that offers a panoramic view of the object. With a 20 Hz and 50 Hz pulse repetition frequency (PRF), it is equivalent to 800 and 2000 azimuthal scanning angles to be completed in a 10 s single-breath-hold period, respectively. This is equivalent to having 100 K and 500 K array elements on the hemisphere, respectively.

The FOV can be adjusted by changing the location of the source transducer. We placed the focus of the single-element ultrasonic transducer at the intersection of these arcs to achieve a ~ 10 cm diameter FOV. The focus can be adjusted from the axis direction to be above the intersection for reducing FOV or below the intersection to further enlarge the FOV. To synchronize the 3D RUST system, we used a pulse wave signal from the functional signal generator to trigger the DAQ modules, the rotation scanner, and the ultrasound source signal generator. The control system was implemented on an Arduino board and a graphical user interface (GUI) in MATLAB 2019.

The positioning of subjects for head, breast, hand, and foot imaging is depicted in Supplementary Fig. 1. For head imaging, subjects assumed a prone position on a height-adjustable bed, and a head support mechanism was employed to stabilize the head. A polyvinylidene chloride film was utilized to provide vertical support and restrict head movement, while also preventing cross-contamination between subjects. To achieve optimal acoustic coupling, a small amount of distilled water was applied to the film, and the bowl was filled with $D_2O$. This configuration minimized light attenuation and ensured effective acoustic coupling between the transducer array and the film. During breast imaging, subjects also assumed a prone position on the bed, with gentle compression of the breast against the polyvinylidene chloride film. Similar to the head imaging setup, the film provided support, minimized vertical movement of the breast, and prevented cross-contamination among subjects. For hand and foot imaging, subjects stood near the system and placed their hand on top of the polyvinylidene chloride film to ensure stability and optimal acoustic coupling. This positioning allowed for accurate imaging while maintaining the necessary stability and contact between the hand and the film surface. Note that the bowl was filled with water for breast, hand, and foot imaging.

**3D RUS-PAT.** The 3D RUST system can be switched to 3D PAT as a hybrid RUS-PAT system by adding laser illumination to the object. We replaced the single source transducer by installing an engineering diffuser (EDC-80, RPC Photonics Inc.) sealed in the lens tube at the intersection of the arc-shaoed detection arrays. The fluctuation of the light intensity distribution across the angular field-of-view is within 10% according to the specifications of the diffuser. We utilized 1064-nm light for illumination with a 20 Hz laser (Litron, PRF: 20 Hz, maximum pulse energy:

~2.5 J). We achieved a similar FOV as RUST of ~ 10 cm because of the large diverging angle of the engineered diffuser. The FOV can be adjusted by using a diffuser with a different diverging angle. For example, we can use another small diverging engineering diffuser (EDC-50, RPC Photonics Inc.) to achieve a 6 cm diameter FOV. For the current setup, the radiant exposure (~ 30 mJ/cm$^2$ at 1064 nm) and fluence rate (~ 300 mW/cm$^2$) are within the ANSI safety limits[41] (100 mJ/cm$^2$ and 1,000 mW/cm$^2$ at 1,064 nm). The mechanical index (MI ~ 0.1) and spatial-peak temporal-average intensity (I$_{SPTA}$ ~ 50 mW/cm$^2$) are within the safety limits defined by U.S. Food and Drug Administration (FDA)'s Electronic Product Radiation Control (EPRC) and the International Electrotechnical Commission (IEC), which require MI < 1.9, and I$_{SPTA}$ < 720 mW/cm$^2$. We first performed the 3D PAT scan for 10 s and then switched the system to perform the 3D RUST scan for another 10 s with the subject in the same position.

**Image reconstruction.** The image reconstruction of 3D RUS-PAT was based on the delay and sum (DAS) and universal back-projection (UBP) algorithms[42,43] implemented in MATLAB and C++. In RUST image reconstruction, we collected the ultrasound scattered signals from each array element and scanning position. Then, we reweighted the received signals based on the density of the array elements after scanning. For each voxel in the reconstructed 3D space, we summed the received signal amplitudes with appropriate delays to obtain the reconstructed image. In PAT image reconstruction, we back-projected the received PA signals at all scanning positions to the 3D space to form a volumetric image. We have also taken into consideration the acceleration and deceleration of the motor at the beginning and end positions during image reconstruction for the coordinates of the synthetic array elements, as illustrated in Supplementary Fig. 13. The voxel size for both modalities is 0.25 × 0.25 × 0.25 mm$^3$. The reconstructed volumetric images were post-processed using depth compensation and Hessian-based Frangi vesselness filtration.

**Imaging protocols**. All human imaging experiments were performed with the relevant guidelines and regulations approved by the Institutional Review Board of the California Institute of Technology (Caltech). We have conducted imaging sessions on a total of seven subjects, including one hemicraniectomy patient's head, two subjects' breasts, two subjects' hands, and two patients' feet. The human experiments were performed in a dedicated imaging room. Written informed consent was obtained from all the participants according to the study protocols.

**Data availability**

The data that support the findings of this study are provided within the paper and its Supplementary Information.

**Code availability**

The reconstruction algorithm and data processing methods can be found in Methods. Owing to a pending patent application, we have opted not to make the computer codes publicly available.

**Acknowledgments**

We thank J. Olick-Gibson for proofreading the manuscript. We thank Guadalupe Corral-Leyva for her assistance with the patients. This work was sponsored by the United States National Institutes of Health (NIH) grants R01 CA282505, U01 EB029823 (BRAIN Initiative), and R35 CA220436 (Outstanding Investigator Award).

**Contributions**

L.V.W., Y.Z., and S.N. designed the study. Y.Z., L.L., S.N., X.T., J.F.Z. and K.M. built and modified the system. Y.Z. developed the RUST image reconstruction algorithm. P.H. developed the PAT image reconstruction algorithm. Y.Z., K.S., and S.N. performed the experiments. Y.J.A. and Y.Z. performed the simulation. Y.Z. and L.V.W. analyzed the RUST data. Y.Z., K.S., Y.L., and L.V.W. analyzed and interpreted the PAT data. J.J.R. and C.Y.L. recruited the participants and interpreted the head data. T.T.W., C.Y.L., and J.J.R. recruited the patients with diabetic foot ulcers. Y.Z. wrote the manuscript with input from all authors. L.V.W. supervised the study and revised the manuscript.

**Competing interests**

L.V.W. has a financial interest in Microphotoacoustics, Inc., CalPACT, LLC, and Union Photoacoustic Technologies, Ltd., which, however, did not support this work.

**Supplementary figures**

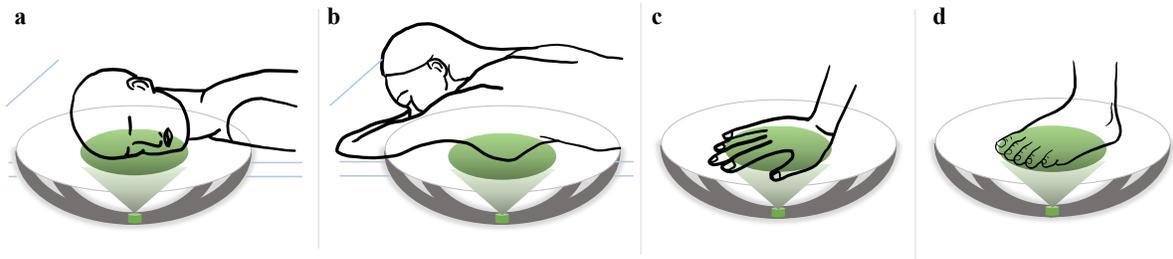

**Supplementary Fig. 1 | Schematic of imaging positions.** The positioning of subjects for head, breast, hand, and foot imaging is depicted in **a**, **b**, **c**, and **d**, respectively. For head and breast imaging, subjects assumed a prone position on a height-adjustable bed, and a support mechanism was employed to stabilize the target. For hand and foot imaging, the subject stood near the system. A polyvinylidene chloride film was utilized to provide vertical support and restrict target movement, while also preventing cross-contamination between subjects. To achieve optimal acoustic coupling, a small amount of distilled water was applied to the film, and the bowl was filled with $D_2O$ or water. This configuration minimized light attenuation and ensured effective acoustic coupling between the transducer array and the film.

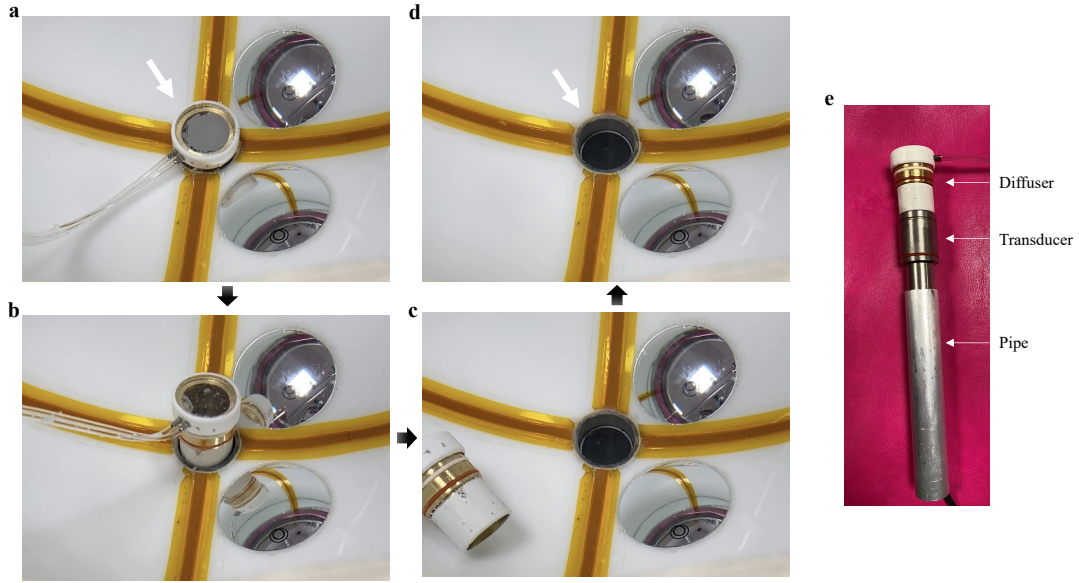

**Supplementary Fig. 2 | Switch between RUST and PAT.** The switch between RUST and PAT ensured that the subject was at the same position for dual-modality data acquisition. **a**, First, the engineered diffuser (white arrow) was mounted at the central hole of the arc-shaped arrays, and we performed the PAT acquisition. **b**, Second, we used a pipe which has the single element ultrasound transducer on the top (**e**) to push out the engineered diffuser and left the ultrasound transducer inside the hole. In this step, we also injected water through the tube to the holder of diffuser to remove the air. **c**, Third, we pulled out the engineered diffuser away from the center of the array. **d**, Finally, the single element transducer (white arrow) was at the center of the array, and we could acquire the RUST image with the subject at the same position with PAT. **e**, Photograph of diffuser, transducer, and the pipe.

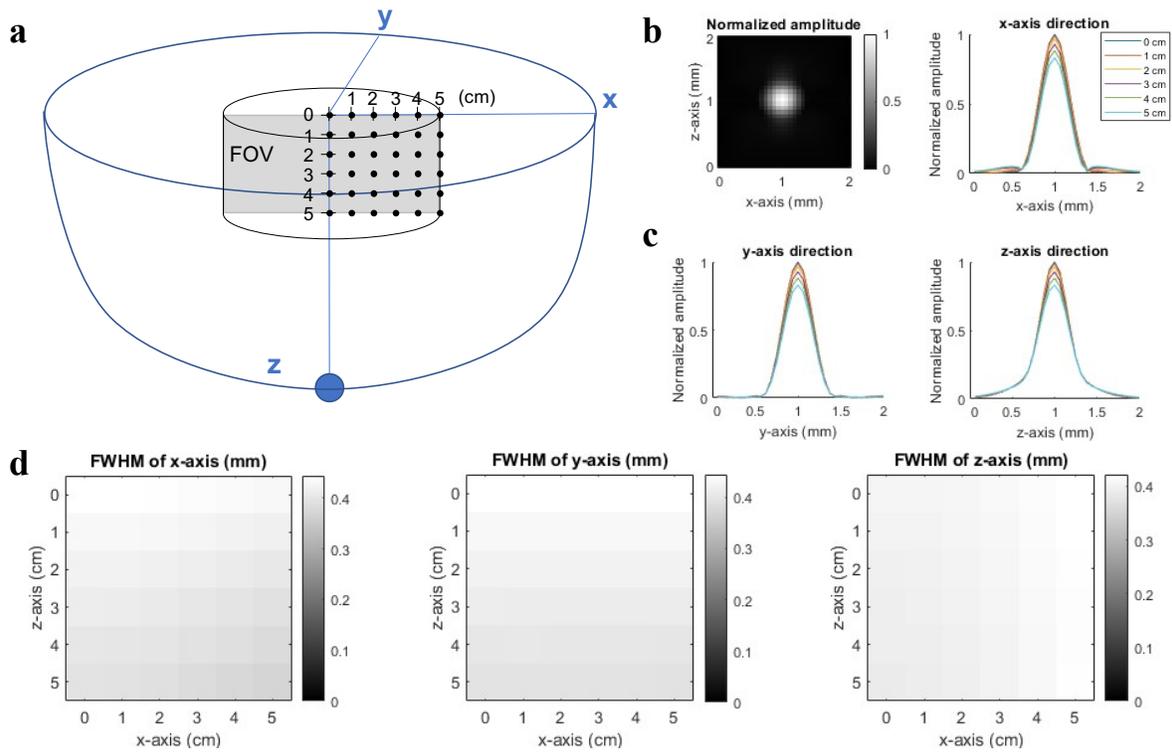

**Supplementary Fig. 3 | Uniformity of spatial resolution for RUST in a numerical simulation. a,** Coordination of RUS-PAT system. Black dots indicate the locations of ultrasound point sources in the simulation. **b,** An example of reconstructed images at a specified position ($x = 0$, $y = 0$, $z = 5$ cm). Note that the axis in the image shows the relative position to the center point. **c,** Amplitude profiles of reconstructed images at a specified line area ($y = 0$ and $z = 5$ cm). Different colors represent each $x$-position of the ultrasound point sources. **d,** Matrix of FWHM values in $x$, $y$, and $z$-axis directions. The maximal percentage changes of FWMH along the $x$, $y$, and $z$ directions are 7.8%, 5.1%, and 3.5%, respectively. FOV: Field-of-view; FWHM: Full width at half maximum.

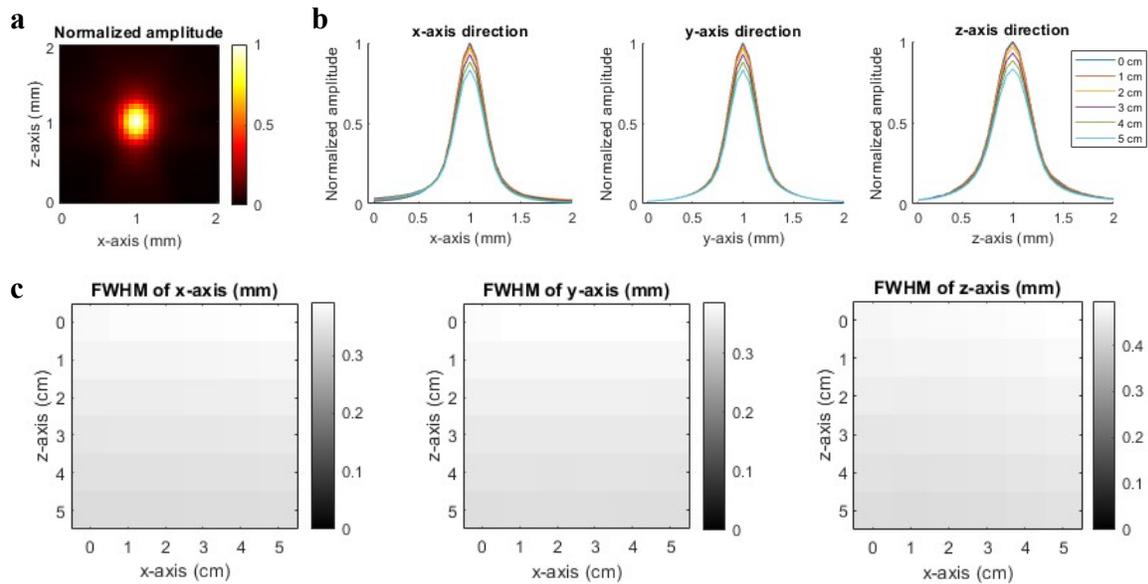

**Supplementary Fig. 4 | Uniformity of spatial resolution for PAT in a numerical simulation. a,** An example of reconstructed images at a specified position ($x = 0$, $y = 0$, $z = 5$ cm). Note that the axis in the image shows the relative position to the center point. **b,** Amplitude profiles of reconstructed images at a specified line area ($y = 0$ and $z = 5$ cm). Different colors represent each $x$-position of the ultrasound point sources. **c,** Matrix of FWHM values in $x$, $y$, and $z$-axis directions. The maximal percentage changes of FWMH along the $x$, $y$, and $z$ directions are 6.3%, 5.7%, and 6.3%, respectively. FWHM: Full width at half maximum.

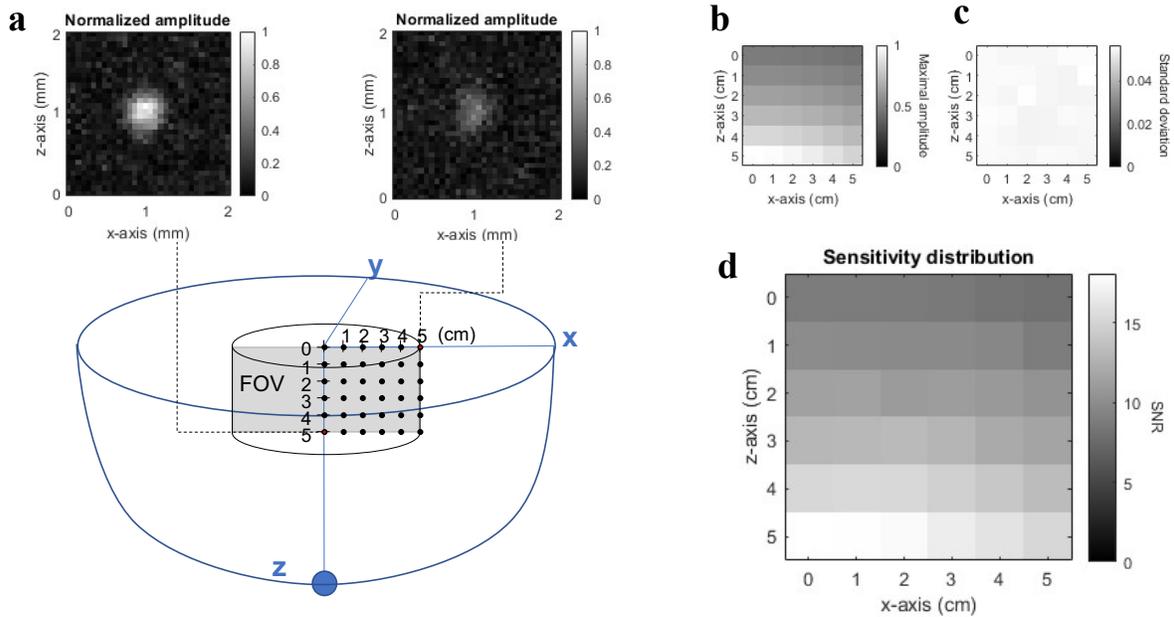

**Supplementary Fig. 5 | Simulation-based quantification of sensitivity distribution of the RUST mode in the FOV.** The signal-to-noise ratio (SNR) was evaluated within a defined region of interest (ROI) on the reconstructed image. The ROI is a 2 mm × 2 mm × 2mm voxel ($32 \times 32 \times 32$ pixels). **a,** An example of the reconstructed RUST image at ($x = 0$, $y = 0$, $z = 5$ cm) and ($x = 5$ cm, $y = 0$, $z = 5$ cm) positions. Note that the axis in the image shows the relative position to the center point. Gaussian noise was added to each receiver array. **b,** Distribution of the maximal amplitude on the reconstructed images with a point source, where there was no noise. **c,** Matrix of the standard deviation of pixel amplitudes within the ROI of the reconstructed image, containing only noise. **d,** Distribution of SNR within the FOV. The maximal percentage change of SNR was 55%.

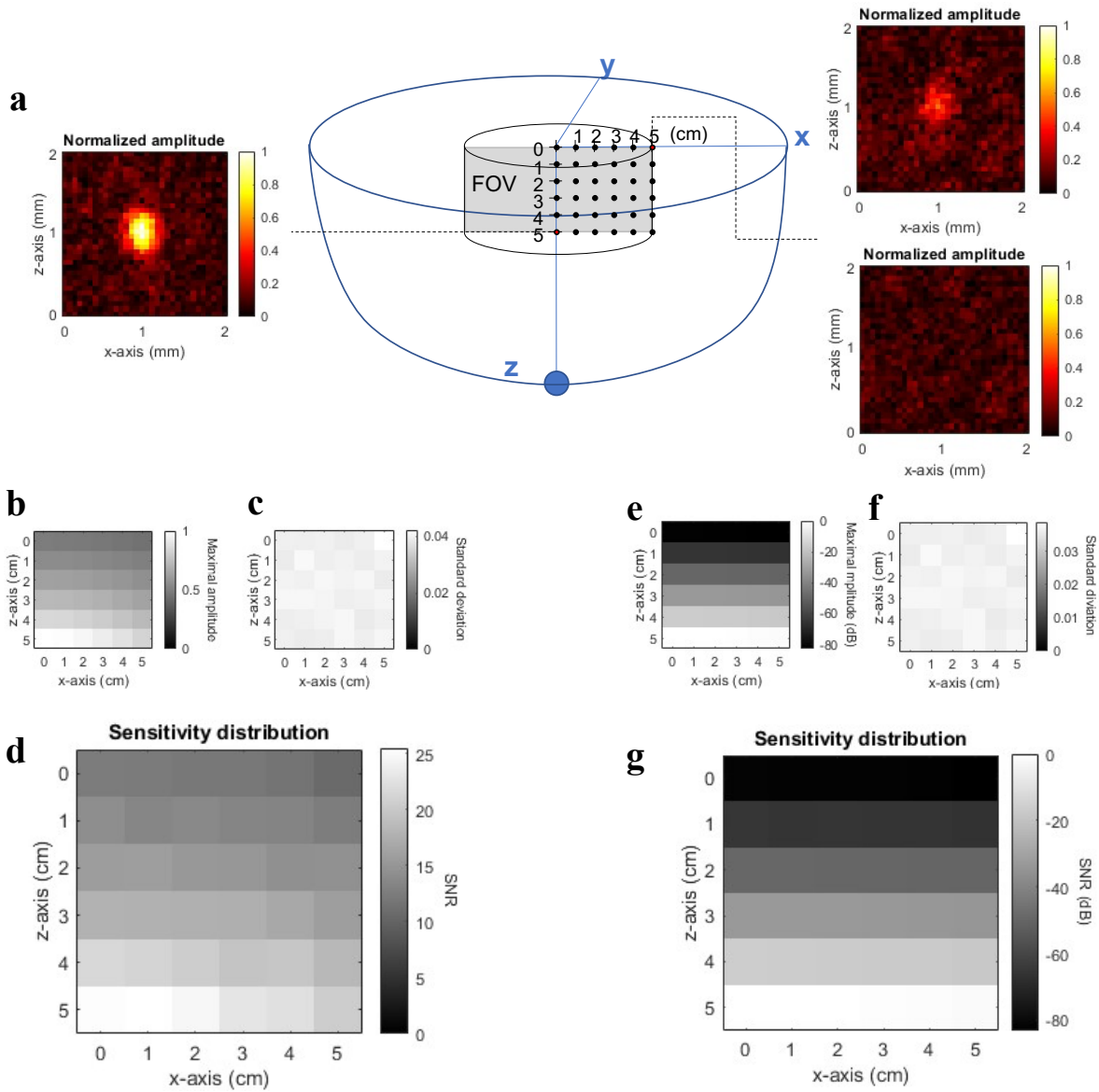

**Supplementary Fig. 6 | Simulation-based quantification of the sensitivity distribution of the PAT mode.** The same method used for evaluating the signal-to-noise ratio (SNR) in the RUST mode was also employed for PAT. **a**, An example of the reconstructed PAT image at specified positions ($x = 0$, $y = 0$, $z = 5$ cm) and ($x = 5$ cm, $y = 0$, $z = 5$ cm). The top right panel represents no light attenuation, while the bottom right panel considers light attenuation in tissue. **b**, Distribution of the maximal amplitude of the reconstructed images utilizing point sources without any additional noise with the consider of acoustic attenuation only. **c**, Matrix of the standard deviation of pixel amplitudes in the ROI of the noise image. **d**. Distribution of SNR in FOV. The maximal percentage change of SNR was 58%. **e**, Distribution of the maximal amplitude of the reconstructed images utilizing point sources without any additional noise with the consideration of both light and acoustic attenuation. **f**. Matrix of the standard deviation of pixel amplitudes in the ROI of the noise image. **g**. The distribution of SNR in the FOV shows a drop of approximately 50 dB with a 4 cm penetration depth.

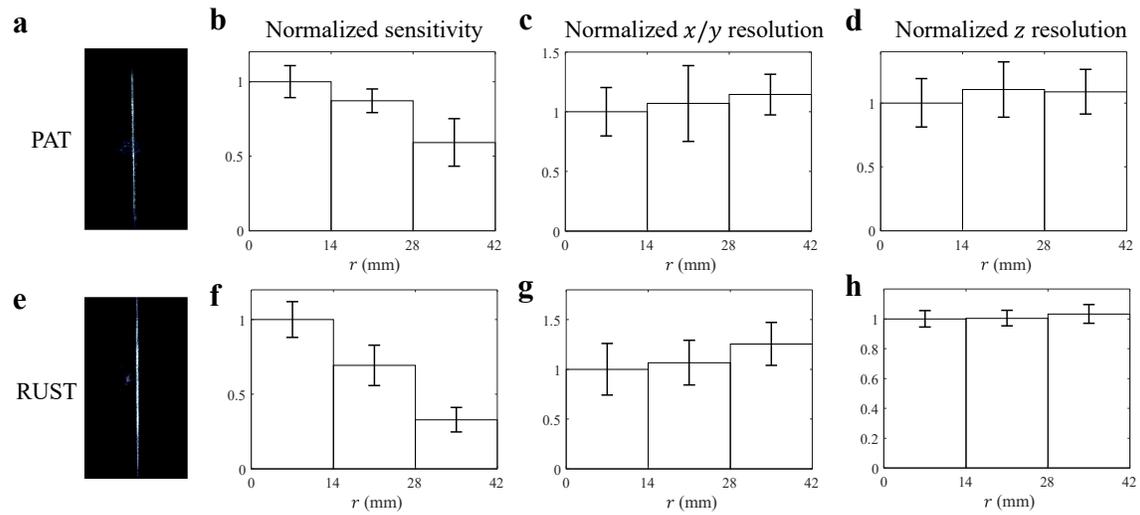

**Supplementary Fig. 7 | Sensitivity and resolution analysis of the RUS-PAT system using a light absorbing and ultrasound scattering line target. a,** PAT image of a line target. **b-d,** Plots of the variation in the normalized sensitivity, $x/y$ resolution, and z resolution of PAT with the radial distance (r) in the $xy$ plane. The line is divided into three segments and the sensitivity and resolutions in each segment are computed as the mean of the respective metric at 56 different uniformly spaced positions (pitch: 0.25 mm) along the segment. The error bars represent the standard deviation of the respective quantity in each segment. **e,** RUST image of a line target. **f-h,** Plots of the variation in the normalized sensitivity, $x/y$ resolution, and $z$ resolution of RUST with the radial distance ($r$) in the $xy$ plane.

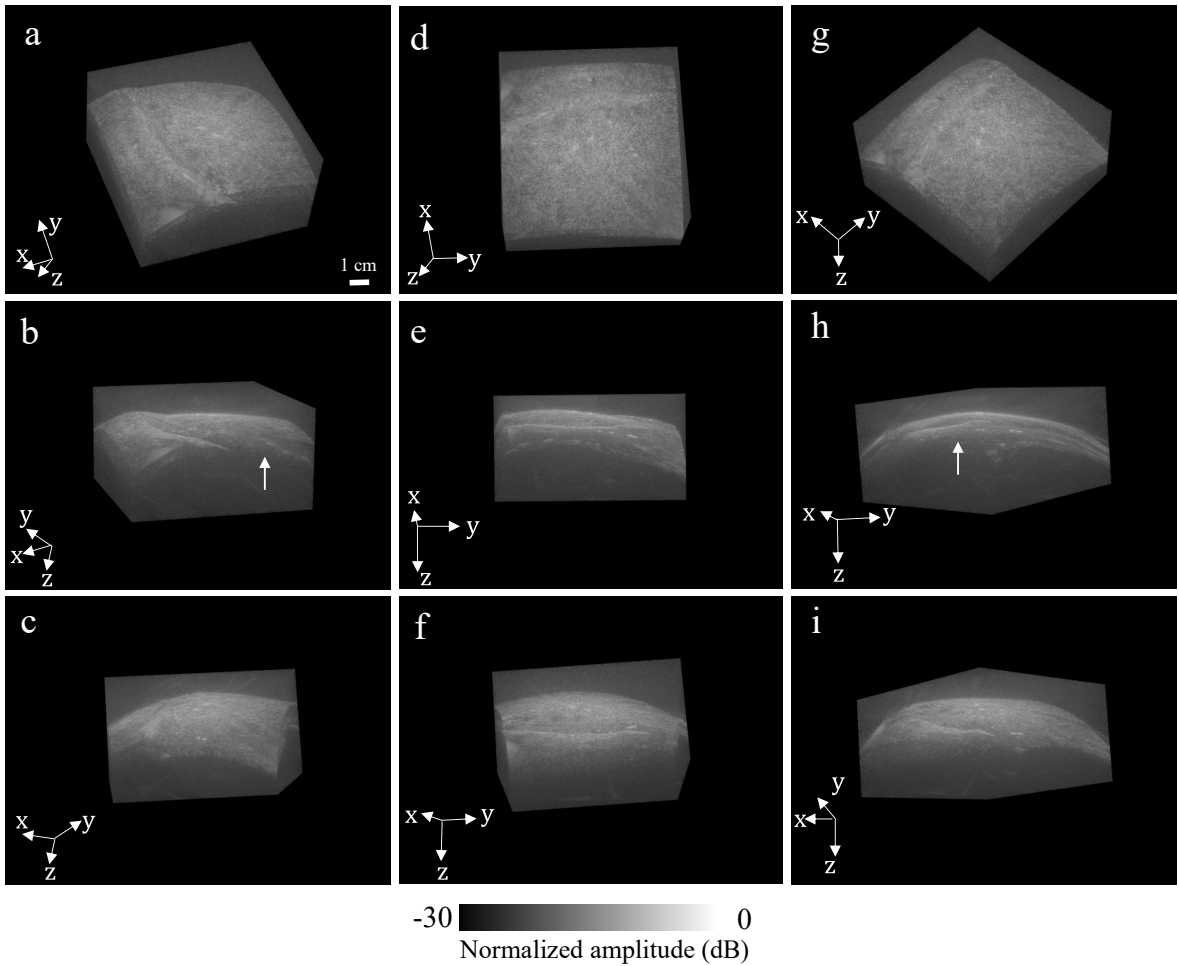

**Supplementary Fig. 8 | 3D RUST of the head. a-i,** Images of a hemicraniectomy patient's head in different views by RUST. White arrows indicate the boundary between the scalp and cortical region.

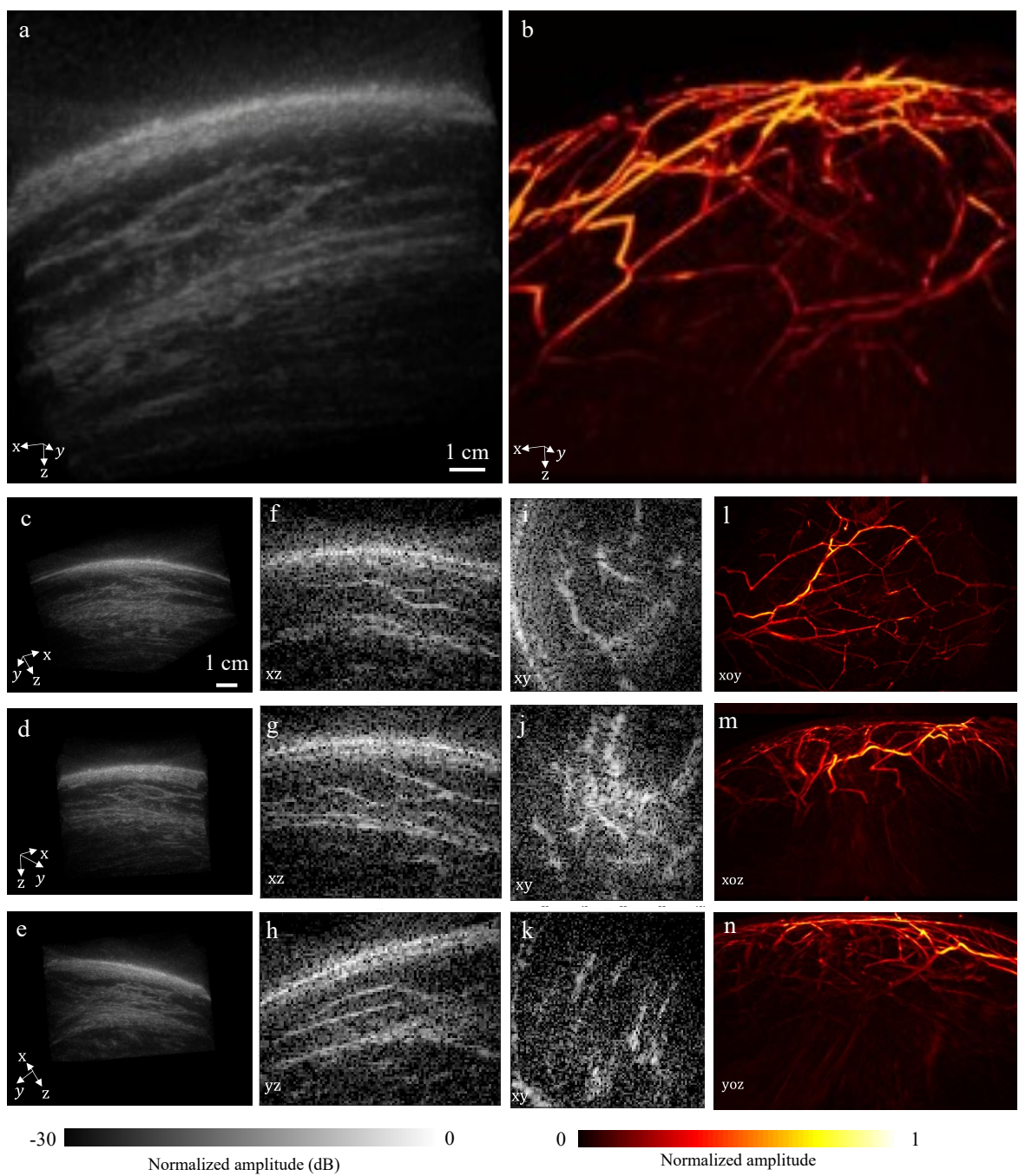

**Supplementary Fig. 9 | 3D RUS-PAT of the breast. a-b,** side-by-side comparison between RUST and PAT MAP images of the breast of a healthy subject. **c-e,** MAP images of the breast by RUST. **f-h,** 2D xz and yz slice-views of the breast images by RUST at $y = 0$ cm, $y = 3$ cm, and $x = 2$ cm, respectively. **i-k,** 2D xy slice-view of the breast images by RUST at depths of 1 cm, 3 cm, and 5 cm, respectively, **l-n,** MAP images of the breast by PAT.

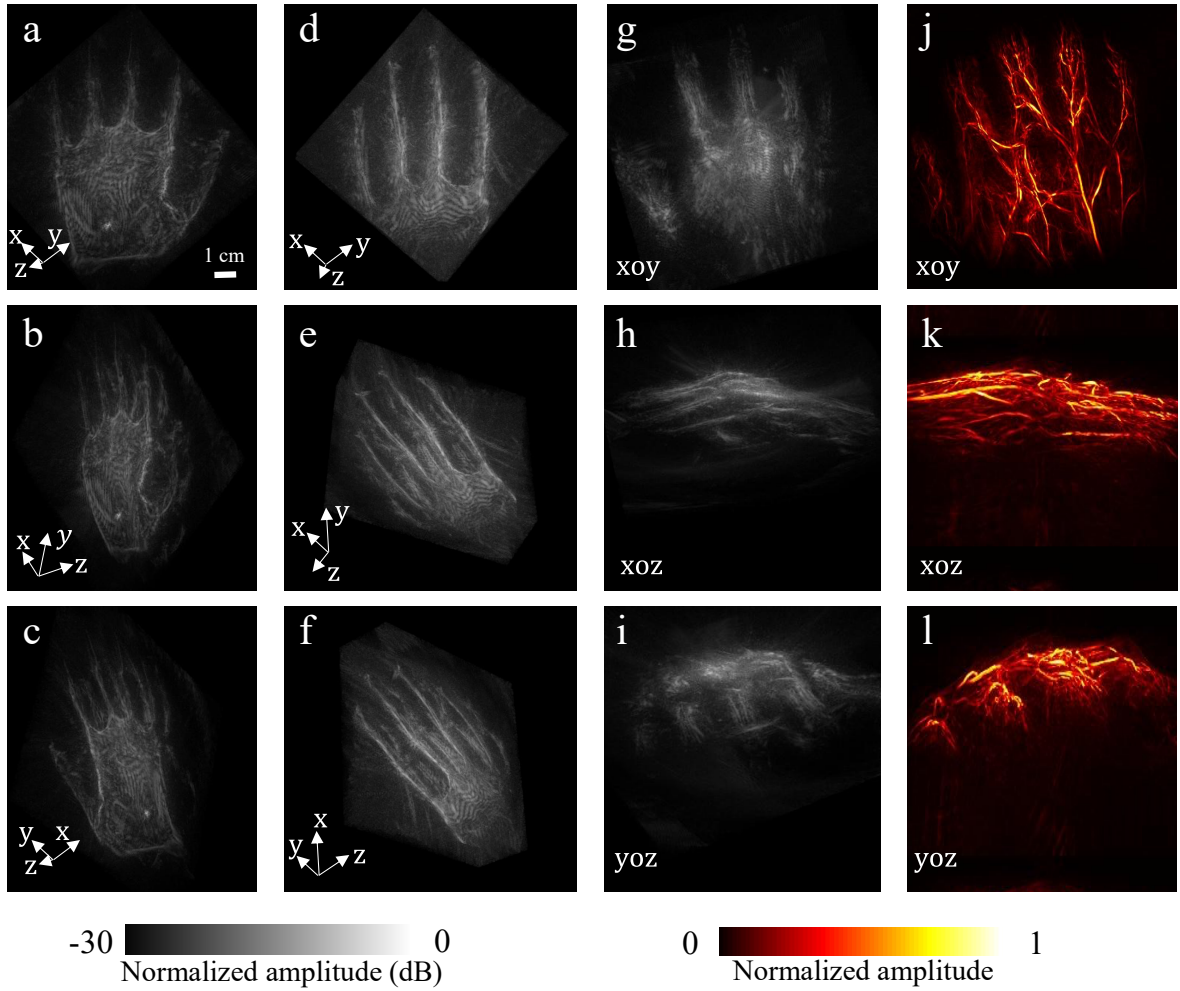

**Supplementary Fig. 10 | 3D RUS-PAT of the hand. a-c,** MAP images of the palm by RUST. **d-f,** MAP images of the fingers by RUST. **g-i,** MAP images of the back of the hand by RUST. **j-l,** MAP images of the back of the hand by PAT.

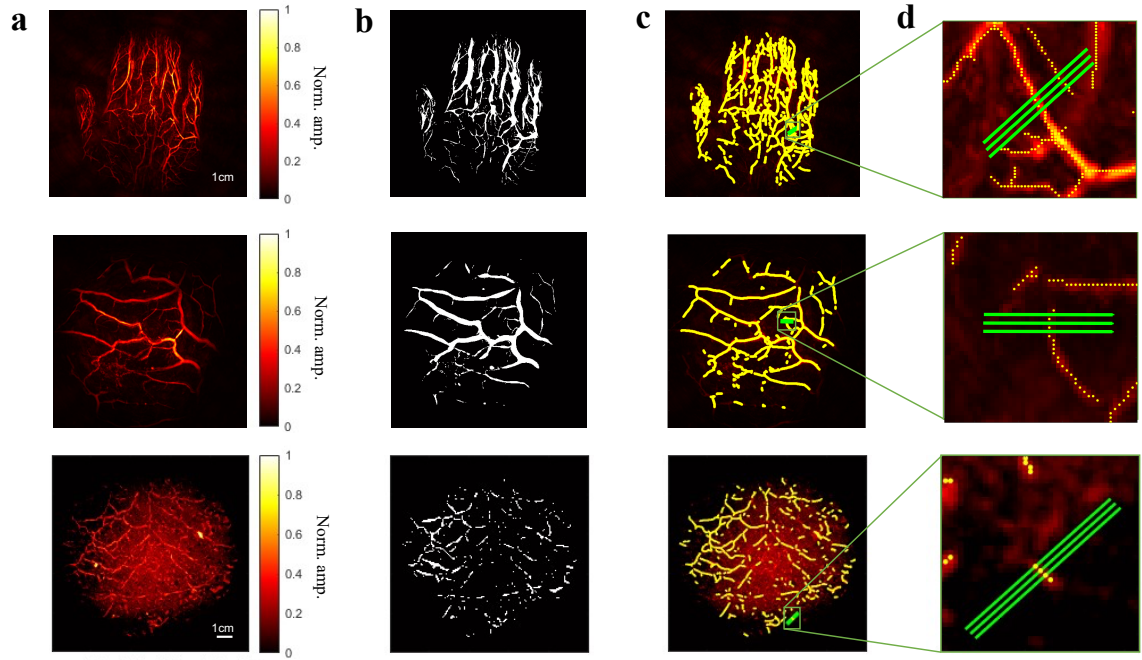

**Supplementary Fig. 11 | Quantification of the sensitivity in terms of blood vessel diameter of the head, breast, and hand PAT images.** Blood vessels on each image were automatically detected by using a Frangi filter and performing vessel segmentation. To consider the sensitivity of the PAT images, SNR was used to set the threshold of images. On each image, the detecting algorithm was iterated by going through 10 different SNR values with an interval of 10% of maximal SNR. Then, the smallest diameters were detected by averaging three diameters of consecutive dots. **a**, Original images of the hand, breast, and head PAT images. **b**, Binary images obtained from vessel segmentation. **c**, Blood vessels (red lines) detected by automatic searching. **d**, Blood vessel segments, with the smallest diameters, that are perpendicular to both the screen and the green lines, along which the vessel diameters are computed. The means and standard deviations of the measured smallest diameters of the blood vessels in the hand, breast, and head PAT images are $0.62 \pm 0.06$ mm (at a depth of 1.0 mm), $0.60 \pm 0.03$ mm (at a depth of 1.0 cm), and $0.61 \pm 0.10$ mm (at a depth of 3.0 mm), respectively.

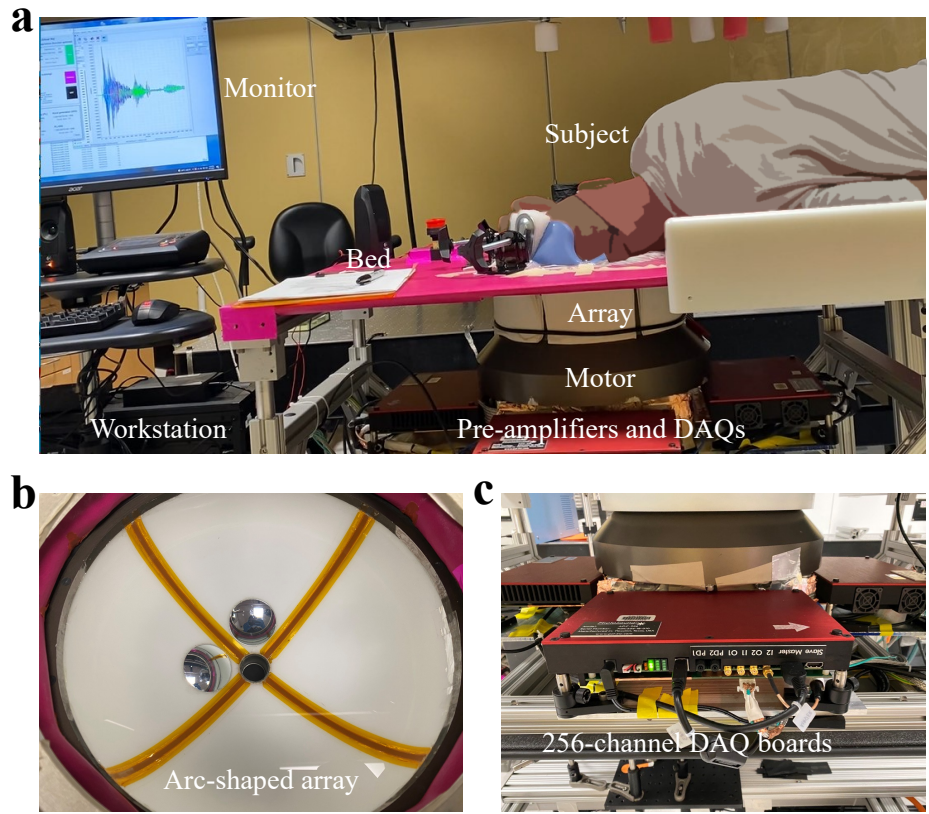

**Supplementary Fig. 12 | Photographs of the system and components. a**, System setup under the bed. The ultrasound transducer array, motor, pre-amplifiers, and DAQs are conveniently installed beneath the bed, allowing the subject to comfortably lie down during system operation. **b**, An example ultrasound transducer array configuration. The array is positioned below the bed to ensure effective coupling with the imaging targets. It comprises four 256-element quarter-ring arrays evenly distributed on a hemispherical housing surface, with a 90-degree pitch between each array. **c**, Acquisition, and amplification of acoustic signals. Four 256-channel DAQ boards (Photosound, Inc.) are integrated on the side of the rotational motor. These boards capture and amplify the acoustic signals, enabling further signal processing.

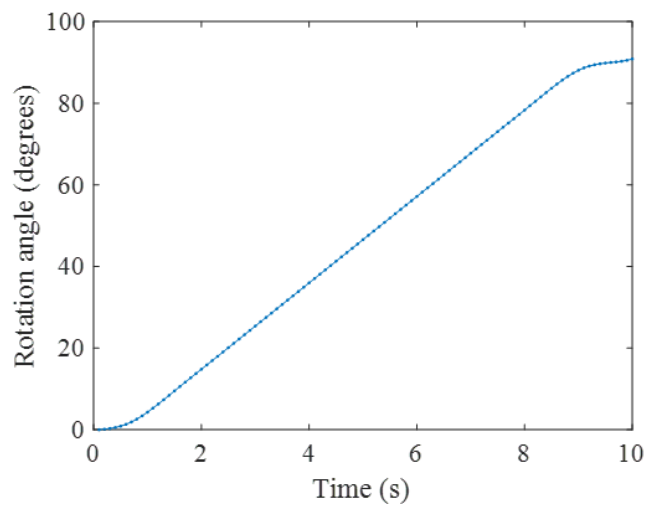

**Supplementary Fig. 13 | Rotation angle versus the scanning time of the motor.** The acceleration and deceleration of the motor are shown at the beginning and end positions, respectively.

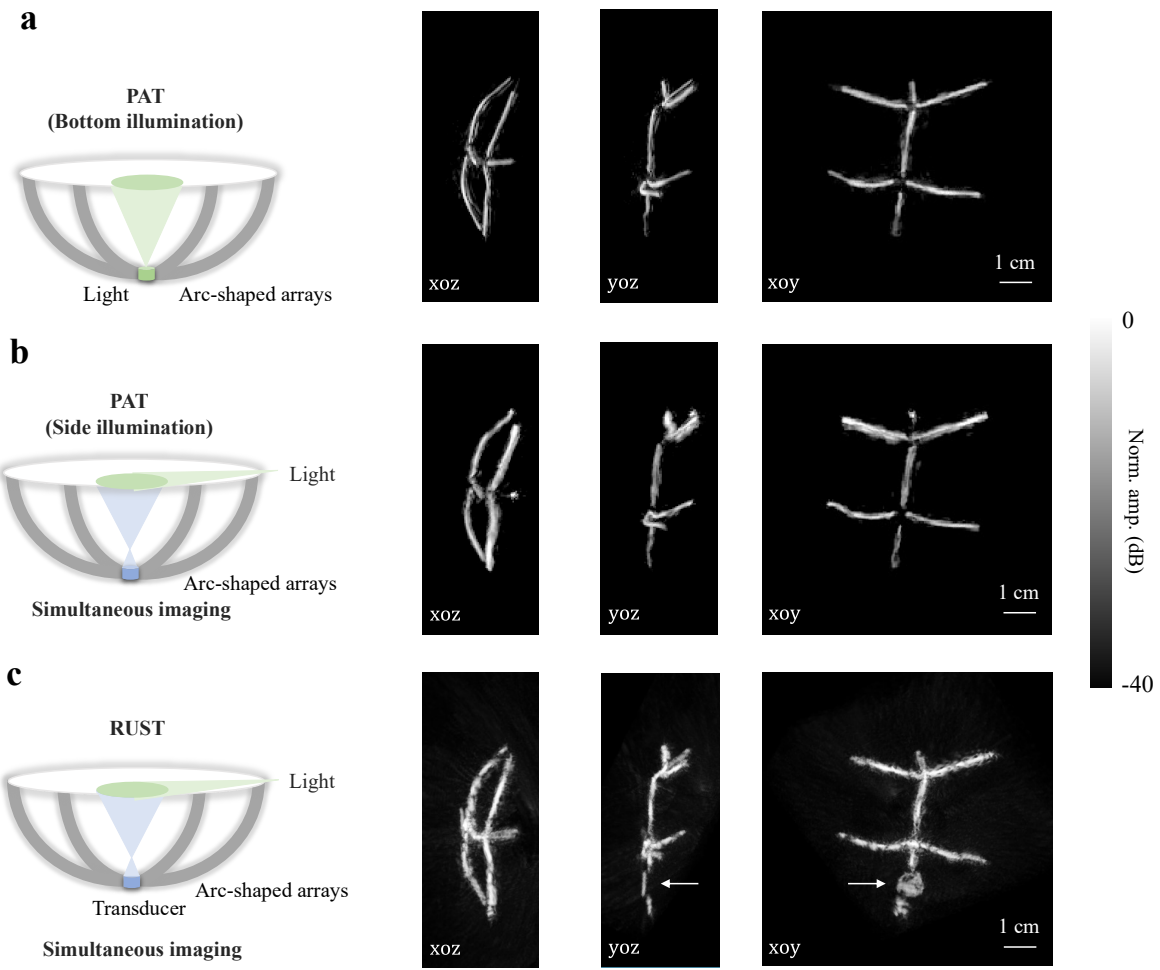

**Supplementary Fig. 14 | Experimental implementation of simultaneous RUST and PAT with side illumination. a,** A schematic and a light-absorbing and ultrasound-scattering phantom PAT images shown in three projected views with light (1064 nm) delivered from the bottom of the array. **b,** A schematic of simultaneous RUST and PAT, along with phantom PAT images displayed in three projected views with light delivered from the side of the array. **c,** A schematic of simultaneous RUST and PAT, along with phantom RUST images shown in three projected views, with the ultrasound transducer positioned at the bottom of the array. The bright spot indicated by the white arrows is due to transparent glue used to hold the target in place.

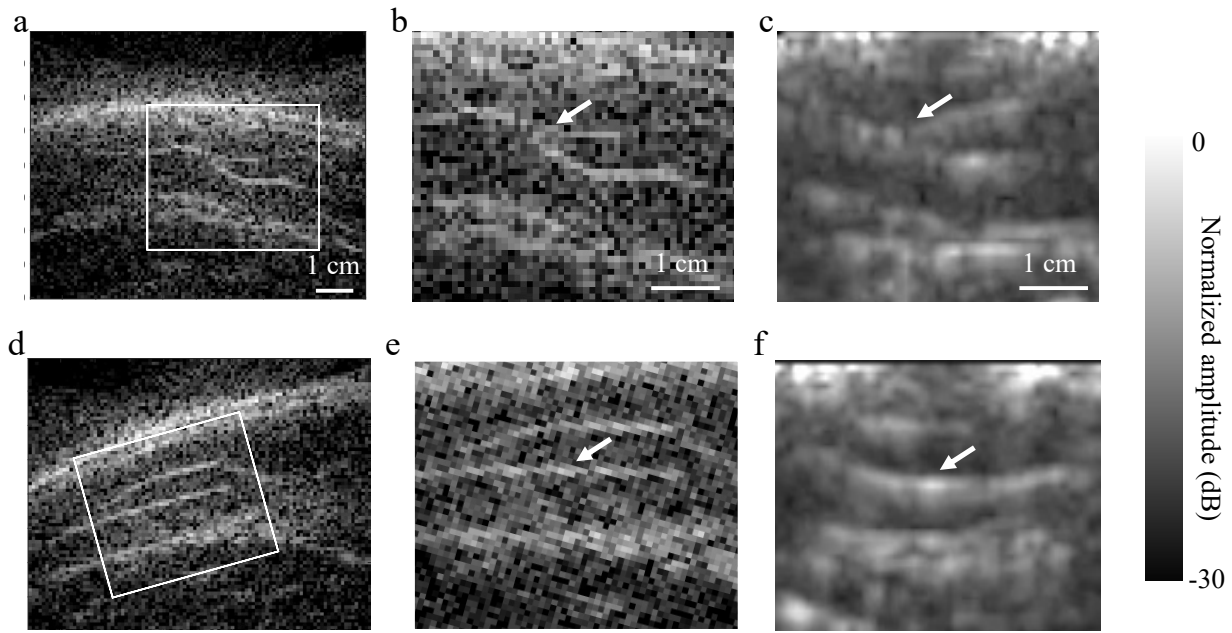

**Supplementary Fig. 15 | *In vivo* breast images of RUST and a linear array at similar planes.** A standard linear array probe was used to image the same breast subject at approximately the same center frequency as the RUS-PAT system. Two comparable planes were selected from the 2D slices of the three-dimensional RUST images and the corresponding linear array images for comparison. **a-c,** 2D $xz$-slice view of breast images acquired by RUST at $y = 0$ cm, showing the original RUST image, a zoomed-in view of the RUST image (white box), and the ultrasound image obtained with the linear array at a similar plane. **d-f,** 2D $yz$-slice view of breast images acquired by RUST at $x = 2$ cm, showing the original RUST image, a zoomed-in view of the RUST image (white box), and the ultrasound image obtained with the linear array at a similar plane. White arrows indicate similar structures in the images obtained by the two methods. The slight discrepancies between **b** and **c**, as well as **e** and **f**, can be attributed to gentle compression of the breast by the linear array probe and positional differences during imaging. RUST: Rotational ultrasound tomography.

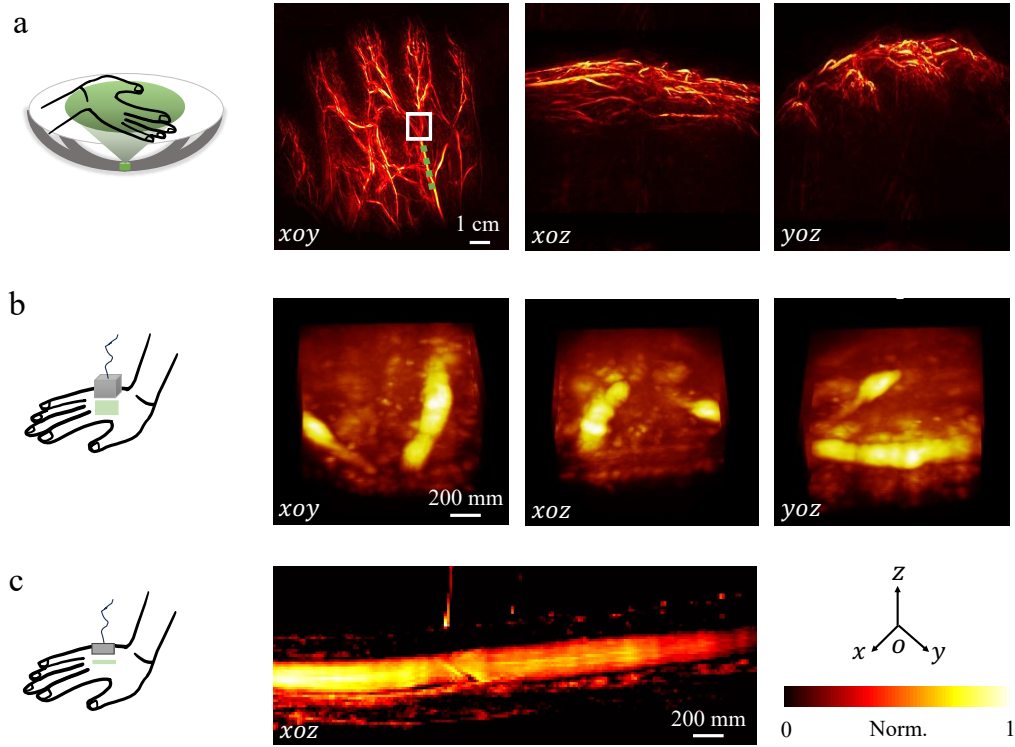

**Supplementary Fig. 16 | *In vivo* hand images obtained using RUS-PAT and state-of-the-art Doppler ultrasound.** The hand images were acquired using the RUS-PAT system and state-of-the-art Doppler ultrasound. **a**, RUS-PAT hand images shown in three projected views: *xoy*, *xoz*, and *yoz*. **b,** Hand images obtained using ultrafast Doppler ultrasound with a matrix array probe (1024 elements, 0.3 mm pitch, 8 MHz center frequency, Vermon, Verasonics, Inc.) in three projected views. The imaged region corresponds to the white box in **a**. **c**, Hand image acquired using ultrafast Doppler ultrasound with a linear array probe (256 elements, 15 MHz center frequency, LZ250, VisualSonics Inc.). The imaged region corresponds to the dashed green line in **a**. RUS-PAT: Rotational ultrasound and photoacoustic tomography.